\documentclass[]{elsarticle}

\makeatletter
\def\ps@pprintTitle{%
 \let\@oddhead\@empty
 \let\@evenhead\@empty
 \def\@oddfoot{}%
 \let\@evenfoot\@oddfoot}
\makeatother

\usepackage{amssymb}
\usepackage{mathtools}
\usepackage{subcaption}
\usepackage[section]{placeins}

\usepackage[numbers]{natbib}
\usepackage{listings}

\usepackage{hyperref}
\usepackage{floatrow}
\usepackage{algorithm}
\usepackage{graphicx}
\usepackage{algpseudocode}
\usepackage{amsmath}
\usepackage[size=footnotesize]{subcaption}
\usepackage[textsize=tiny]{todonotes}
\graphicspath{{figures/}}
\usepackage{cleveref}
\crefname{appendix}{}{}
\usepackage{soul}
\usepackage{geometry}
\usepackage{algorithmicx}
\usepackage{booktabs}
\usepackage{longtable}
\usepackage{siunitx}
\usepackage{stmaryrd}
\usepackage{algpseudocode}
\usepackage{adjustbox}
\usepackage{color, colortbl}
\usepackage{multirow}
\usepackage{xcolor}
\usepackage{ulem}
\usepackage{svg-extract}
\usepackage{svg}

\newcommand{\Sf}{\mathbf{S}_f}

\let\textv\v
\renewcommand{\v}{\TextOrMath{\textv}{\mathbf{v}}}
\newcommand{\g}{\mathbf{g}}
\newcommand{\x}{\mathbf{x}}
\newcommand{\n}{\mathbf{n}}
\renewcommand{\H}{\mathbf{H}}

\newcommand{\OF}{OpenFOAM\textsuperscript{\textregistered}\ }

\renewcommand{\O}{\mathbf{x}_{O_f}}
\newcommand{\N}{\mathbf{x}_{N_f}}
\newcommand{\df}{\mathbf{d}_f}
\newcommand{\dfhat}{\hat{\mathbf{d}}_f}

\newcommand{\Sfnorth}{\Sf^{\|}}
\newcommand{\Sforth}{\Sf^{\perp}}

\colorlet{Reviewer1}{violet}
\colorlet{Reviewer2}{red}

\allowdisplaybreaks

\begin{document}

\frenchspacing

\begin{frontmatter}


\author[add1]{Jun Liu}
\ead{liu@mma.tu-darmstadt.de}
\author[add2]{Tobias Tolle}
\ead{tobias.tolle@de.bosch.com}
\author[add1]{Tomislav Mari\'{c}\corref{corr}}
\cortext[corr]{Corresponding author}
\ead{maric@mma.tu-darmstadt.de}

\address[add1]{Mathematical Modeling and Analysis, Technische Universit\"{a}t Darmstadt, Germany}
\address[add2]{Bosch Research, Robert Bosch GmbH; research conducted at MMA, Technische Universit\"{a}t Darmstadt}



\title{A residual-based non-orthogonality correction for force-balanced unstructured Volume-of-Fluid methods}


\begin{abstract}

Non-orthogonality errors in unstructured Finite Volume methods for simulating incompressible two-phase flows may break the force-balanced discretization. We show that applying the same explicit non-orthogonality correction for all gradient terms in the context of segregated solution algorithms is not sufficient to achieve force balance. To ensure force balance, we introduce a straightforward and deterministic residual-based control of the non-orthogonality correction, which removes the number of non-orthogonality corrections as a free parameter from the simulation. Our method is directly applicable to different unstructured finite-volume two-phase flow simulation methods as long as they discretize the one-field formulation of incompressible two-phase Navier-Stokes equations. We demonstrate force balance for the surface tension force and the gravity force near linear solver tolerance for an algebraic and a geometric Volume-of-Fluid method using the stationary droplet and stationary water column verification cases on polyhedral unstructured meshes with varying levels of non-orthogonality.
\end{abstract}




\end{frontmatter}

\section{Introduction}
\label{sec:intro}

Numerical methods for simulating two-phase flows must ensure a balance of forces acting on the fluid interface on the discrete level. An imbalanced discretization makes it impossible in some cases to achieve a steady-state interface shape, e.g. for a canonical case of a spherical droplet suspended in air without the influence of gravity, or a stationary droplet that is wetting a surface. 

We focus on the force-balanced discretization in the unstructured Finite Volume Method \citep{Jasak1996, Maric2014,Darwish2017}, because of its high degree of volume conservation and its ability to discretize boundary conditions at geometrically complex domain boundaries with second-order accuracy. We implement our approach and verify it for the unstructured geometrical Volume-of-Fluid (VOF) method \citep{Scheufler2019} and the unstructured algebraic VOF method \citep{Deshpande2012}.

In \cref{sec:math_model} we discuss the governing equations, and in \cref{sec:methodology} we describe their force-balanced discretization and their solution algorithm. We demonstrate in \cref{sec:methodology} that the same principle of error cancellation in the structured force-balanced discretization \citep{Popinet2018} intuitively extends to unstructured meshes, contrary to recent findings in \citep{Huang2023}, under the condition that the pressure and velocity equation solution converges. In \cref{sec:state-of-the-art}, with the knowledge about the mathematical model and the unstructured Finite-Volume and VOF discretization from \cref{sec:math_model,sec:methodology}, respectively, we compare our approach with state-of-the-art methods.
\section{Mathemathical model}
\label{sec:math_model}

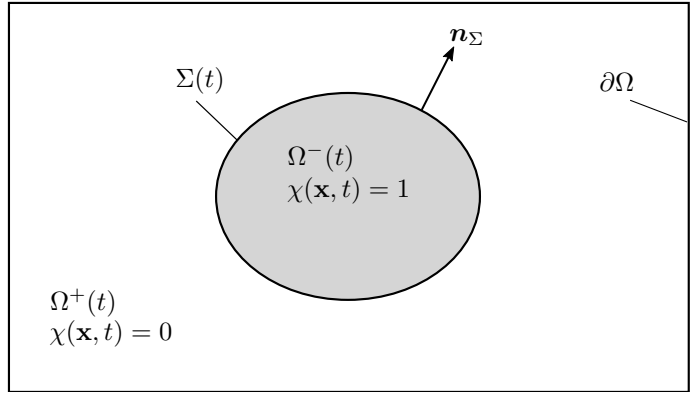
\begin{figure}[!htb]
    \centering
    \def\svgwidth{0.6\textwidth}
\begingroup%
  \makeatletter%
  \providecommand\color[2][]{%
    \errmessage{(Inkscape) Color is used for the text in Inkscape, but the package 'color.sty' is not loaded}%
    \renewcommand\color[2][]{}%
  }%
  \providecommand\transparent[1]{%
    \errmessage{(Inkscape) Transparency is used (non-zero) for the text in Inkscape, but the package 'transparent.sty' is not loaded}%
    \renewcommand\transparent[1]{}%
  }%
  \providecommand\rotatebox[2]{#2}%
  \newcommand*\fsize{\dimexpr\f@size pt\relax}%
  \newcommand*\lineheight[1]{\fontsize{\fsize}{#1\fsize}\selectfont}%
  \ifx\svgwidth\undefined%
    \setlength{\unitlength}{475.33460242bp}%
    \ifx\svgscale\undefined%
      \relax%
    \else%
      \setlength{\unitlength}{\unitlength * \real{\svgscale}}%
    \fi%
  \else%
    \setlength{\unitlength}{\svgwidth}%
  \fi%
  \global\let\svgwidth\undefined%
  \global\let\svgscale\undefined%
  \makeatother%
  \begin{picture}(1,0.57363443)%
    \lineheight{1}%
    \setlength\tabcolsep{0pt}%
    \put(0,0){\includegraphics[width=\unitlength,page=1]{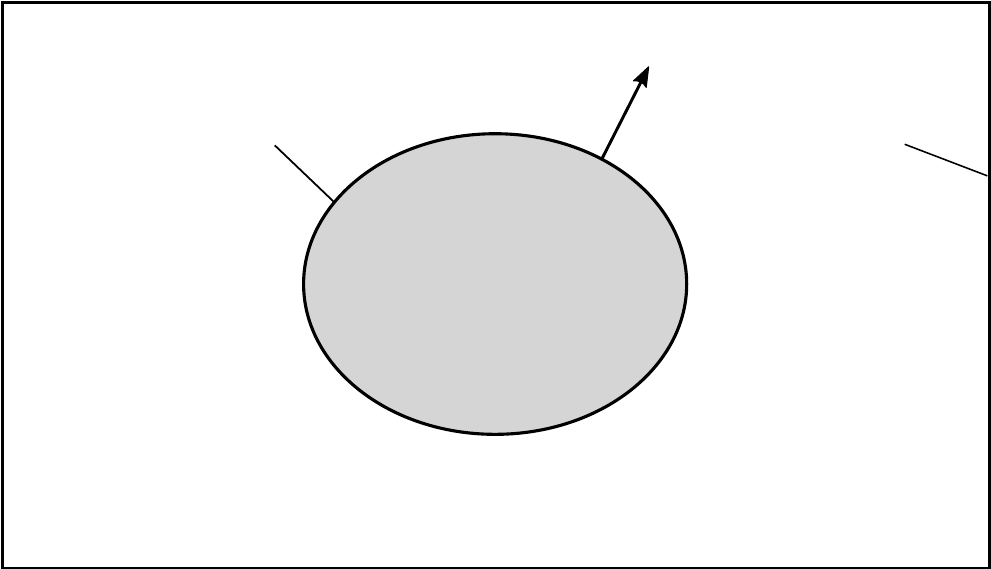}}%
    \put(0.41135156,0.33339009){\color[rgb]{0,0,0}\makebox(0,0)[lt]{\smash{\begin{tabular}[t]{l}$\Omega^-(t)$\end{tabular}}}}%
    \put(0.41067232,0.28531297){\color[rgb]{0,0,0}\makebox(0,0)[lt]{\smash{\begin{tabular}[t]{l}$\chi(\mathbf{x}, t) = 1$\end{tabular}}}}%
    \put(0.06215877,0.07651644){\color[rgb]{0,0,0}\makebox(0,0)[lt]{\smash{\begin{tabular}[t]{l}$\chi(\mathbf{x}, t) = 0$\end{tabular}}}}%
    \put(0.8689053,0.44140486){\color[rgb]{0,0,0}\makebox(0,0)[lt]{\smash{\begin{tabular}[t]{l}$\partial \Omega$\end{tabular}}}}%
    \put(0.64858374,0.51490576){\color[rgb]{0,0,0}\makebox(0,0)[lt]{\smash{\begin{tabular}[t]{l}$\boldsymbol{n}_{\Sigma}$\end{tabular}}}}%
    \put(0.06311562,0.12014973){\color[rgb]{0,0,0}\makebox(0,0)[lt]{\smash{\begin{tabular}[t]{l}$\Omega^+(t)$\end{tabular}}}}%
    \put(0.24701592,0.4468515){\color[rgb]{0,0,0}\makebox(0,0)[lt]{\smash{\begin{tabular}[t]{l}$\Sigma(t)$\end{tabular}}}}%
  \end{picture}%
\endgroup%
\notag
    \caption{The domain $\Omega$, split by the fluid interface $\Sigma(t)$ into two sub-domains $\Omega^\pm$.}
    \label{fig:drop-gas}
\end{figure}

\Cref{fig:drop-gas} illustrates an incompressible two-phase flow system without phase change. The flow domain $\Omega\subset \mathbb{R}^3$ with a boundary $\partial \Omega$ is partitioned into two subdomains $\Omega^+(t)$ and $\Omega^-(t)$, occupied by two different fluid phases. The boundary between $\Omega^\pm(t)$, referred to as the fluid interface, is denoted by $\Sigma(t)$ and has the normal $\mathbf{n}_{\Sigma}$ pointing, say, outwards of the domain $\Omega^-(t)$. 
%
A phase indicator function 
\begin{equation}
  \chi(\x,t) := 
    \begin{cases}
      1, & \x \in \Omega^-(t), \\ 
      0, & \x \in \Omega^+(t), 
    \end{cases}
  \label{eq:indicator}
\end{equation}
indicates $\Omega^\pm(t)$ and formulates single-field density and dynamic viscosities as
\begin{equation}
    \rho(\x,t) = (\rho^-  - \rho^+) \chi(\x,t) + \rho^+, \label{eq:rhoindicator}
\end{equation}
\begin{equation}
    \mu(\x,t) = (\mu^-  - \mu^+) \chi(\x,t) + \mu^+, \label{eq:nuindicator} 
\end{equation}
where $\rho^\pm$ and $\mu^\pm$ are the constant densities and dynamic viscosities in $\Omega^\pm(t)$. The one-fluid density and dynamic viscosity are used in the Navier-Stokes equations without phase change, namely
\begin{align}
    \nabla\cdot\v &= 0 \label{eq:volume-transport}, \\
    \partial_t(\rho \v)+\nabla\cdot(\rho \v \otimes \v)& = -\nabla P + \rho\mathbf{g}
    + \nabla \cdot\left(\mu\left(\nabla\v + (\nabla\v)^T\right)\right)  + \mathbf{f}_\Sigma. 
    \label{eq:momentum-orig}
\end{align}

The dynamic pressure $p$ in the momentum equation \ref{eq:momentum-orig} is a result from splitting the total pressure $P$ on the r.h.s. of the momentum conservation equation into the dynamic pressure $p$ and the hydrostatic pressure
\begin{equation}
     P = p + \rho \mathbf{g} \cdot \mathbf{x},
     \label{eq:definition_pd}
\end{equation}
with $\x$ as the position vector giving $\nabla \x = I$, replacing $-\nabla P$ on the r.h.s. of the momentum conservation equation with
\begin{equation}
        -\nabla P = -\nabla p -  (\mathbf{g}\cdot\mathbf{x}) \nabla \rho
        - \rho\mathbf{g},
\end{equation}
canceling out $\rho\mathbf{g}$ in \cref{eq:momentum-orig}, thus avoiding the use of hydrostatic depth (height) in boundary conditions for the pressure equation, and setting the stage for the balance between the pressure gradient and the gravitational force in the discretization.
The surface tension force $\mathbf{f}_\Sigma$ is exerted only on the interface and modeled by the continuum surface force (CSF) model \citep{Brackbill1992}, as
\begin{equation}
    \mathbf{f}_\Sigma = \sigma \kappa \mathbf{n}_\Sigma\delta_\Sigma \approx -\sigma \kappa \nabla \chi, 
    \label{eq:csf-model}
\end{equation}
where $\sigma$ is the constant surface tension coefficient, $\kappa$ is twice the local mean curvature of the interface $\Sigma(t)$, and $\delta_\Sigma$ is the interface Dirac distribution. The CSF model and its force-balanced discretization played a pivotal role in
two-phase flow simulations and the reader is directed to \citep{Popinet2018} for an informative review of CSF and other surface tension force models. While other models may also be force-balanced, we rely on the CSF model because it permits us to derive the force balance in discretized equations.

The resulting single-field momentum conservation states
\begin{equation}
    \partial_t(\rho \v)+\nabla\cdot(\rho \v \otimes \v) = -\nabla p -  (\mathbf{g}\cdot\mathbf{x}) \nabla \rho + \nabla\cdot\left(\mu\left(\nabla\v + (\nabla\v)^T\right)\right) - \sigma \kappa \nabla \chi. 
     \label{eq:momentum-transport}
\end{equation}

The \cref{eq:indicator,eq:rhoindicator,eq:nuindicator,eq:volume-transport,eq:momentum-orig,eq:momentum-transport} describe mass (volume) and momentum conservation with the help of the phase indicator. Solving equations \cref{eq:indicator,eq:rhoindicator,eq:nuindicator,eq:volume-transport,eq:momentum-orig,eq:momentum-transport} requires the knowledge of $\chi(\x,t)$, coupling volume and momentum conservation
with the phase indicator, advected with the two-phase flow velocity $\v(\x,t)$ \citep{Liu2023}. Modeling and, subsequently,
discretizing $\chi(\x,t)$ or $\Sigma(t)$ is where numerous two-phase flow simulation methods that use the single-field
formulation of Navier-Stokes equations \cref{eq:indicator,eq:rhoindicator,eq:nuindicator,eq:volume-transport,eq:momentum-transport} differ from each other.

We focus in the next sections on the Unstructured Finite Volume Method (UFVM) for discretizing equations  \cref{eq:indicator,eq:rhoindicator,eq:nuindicator,eq:volume-transport,eq:momentum-transport} in geometrically complex domains, leading to so-called non-orthogonality errors, that we correct efficiently with a deterministic iterative solution algorithm. Our approach applies to any two-phase flow simulation method that uses \cref{eq:indicator,eq:rhoindicator,eq:nuindicator,eq:volume-transport,eq:momentum-transport}, irrespective of the way it numerically tracks $\chi(\x,t)$.

\section{Numerical methodology}\label{sec:methodology}

We investigate the impact of the UFVM on the balance of forces at the fluid interface when the UFVM meshes are non-orthogonal and develop a highly accurate, computationally efficient and deterministic stopping criterion for the iterative non-orthogonality correction in the UFVM. Handling non-orthogonality in UFVM is crucial for simulating two-phase flows in geometrically complex domains whose UFV discretization results in a larger degree of non-orthogonality.

We implement our non-orthogonality correction into the unstructured geometrical VOF method \citep{Maric2020}, specifically the plicRDF-isoAdvector method \citep{Scheufler2019}. Our research software \citep{nonOrthogCodes2023,nonOrthogRepository} and research data \citep{nonOrthogData2023} are publicly available.

\subsection{The unstructured geometrical Volume-Of-Fluid Method}
The \textit{volume fraction} is defined as
\begin{equation}
    \alpha_c(t) := \frac{1}{|\Omega_c|}\int_{\Omega_c}\chi(\x, t) \, dV,
    \label{eq:alphadef}
\end{equation}
a fill-level of the phase $\Omega^-(t)$ in a fixed control volume $\Omega_c$. Applying the Reynolds transport theorem to the conservation of the phase indicated by the phase-indicator function $\chi(\x,t)$ (given by \cref{eq:indicator}) within the same $\Omega_c$ with velocity field $\v$, results in 
\begin{equation}
        \partial_t \int_{\Omega_c} \chi(\x, t) \, dV = - \int_{\partial \Omega_c} \chi(\x, t) \v \cdot \n \, dS.
        \label{eq:chitransp}
\end{equation}
Dividing the transport equation \cref{eq:chitransp} by cell volume $|\Omega_c|$ 
and substituting \cref{eq:alphadef}, the flux-based VOF advection equation can be obtained as follows
\begin{equation}
    \partial_t \alpha_c(t) = - \frac{1}{|\Omega_c|}\int_{\partial \Omega_c} \chi(\x, t) \v \cdot \n \, dS.
    \label{eq:alphatransp}
\end{equation}
Integrating \cref{eq:alphatransp} over the temporal interval (\textit{time step}) $[t^n,t^{n+1}]$ with the shorthand notation $\phi^n:=\phi(t^n)$ results in
\begin{equation}
    \alpha_c^{n+1} = \alpha_c^{n} - \frac{1}{|\Omega_c|}\int_{\partial\Omega_c}\int_{t^n}^{t^{n+1}} \chi(\x, t) \v \cdot\n dt dS,
    \label{eq:alphadiscrfromchi}
\end{equation}
which is still an exact equation (\citep{Maric2020,Maric2018}) and expressing the integral on the r.h.s. as $|V_f^{\alpha}|$ - the fluxed phase-specific volume - results in
\begin{equation}
    \alpha_c^{n+1} = \alpha_c^{n} - \frac{1}{|\Omega_c|}\sum_{f \in F_c } |V_f^{\alpha}|_S,
    \label{eq:alphadiscr}
\end{equation}
for the domain $\Omega$ discretized by $|C|$ non-overlapping control volumes $\Omega:=\cup_{c\in C}\Omega_c$. Each finite volume $\Omega_c$ is bounded by a number, $|F_c|$, of non-planar surfaces (faces) $S_f$, i.e. $\partial \Omega_c=\cup_{f \in F_c}S_f$. The singed magnitude of the volume of the phase $\Omega^-(t)$ fluxed over $S_f$, called phase-specific volume, is defined as
\begin{equation}
    |V_f^{\alpha}|_S:=sgn(F_f)|V_f^{\alpha}|,
\end{equation}
where the volumetric flux over the face $S_f$ is given as
\begin{equation}
    F_f:=\int_{S_f}\v\cdot\n dS,
    \label{eq:fluxdef}
\end{equation}
and 
\begin{equation}
    |V_f^{\alpha}|:= \left|\int_{S_f}\int_{t^n}^{t^{n+1}}\chi\v\cdot\n dtdS \right|
\end{equation}
is calculated geometrically \citep{Maric2020}. Flux-based methods are locally and globally highly
volume conservative and, therefore, a preferred choice for incompressible two-phase flows in geometrically complex domains. We implement our approach to the non-orthogonality correction in the plicRDF-isoAdvector unstructured geometrical VOF method in OpenFOAM \citep{Scheufler2019}, which provides us with $\alpha_c^{n+1}$ by geometrically approximating $|V_f^\alpha|_s$ from \cref{eq:alphadiscr}. To demonstrate that our non-orthogonality correction method applies to other methods that discretize single-field Navier-Stokes equations for incompressible two-phase flows, we also implemented our method in the algebraic VOF method in OpenFOAM \citep{Deshpande2012}.

\subsection{Solution algorithm}

Discretizing \cref{eq:volume-transport,eq:momentum-transport} with the implicit Unstructured Finite Volume Method (UFVM) results in an algebraic equation system 
\begin{equation}
    \sum_{f\in F_c} F_f^{n+1}=0,
    \label{eq:voldiscr}
\end{equation}
\begin{equation}
        a_c \v_c^{n+1} + \sum_{n \in N} a_n \v_n^{n+1}  = 
        - (\nabla p)_c^{n+1} - ((\g \cdot \x)\nabla \rho)_c^{n+1}
        - \sigma_c (\kappa\nabla \alpha)_c^{n+1} + \mathbf{S}_c(\v_c^n),
        \label{eq:momdiscrcoeff}
\end{equation}
solved together with the UGVOF equation \cref{eq:alphadiscr}. In \cref{eq:momdiscrcoeff}, $a_c$ is the linear equation system coefficient of the finite volume $\Omega_c$, $N$ is the set of finite volumes that are face-adjacent to $\Omega_c$, and $\mathbf{S}_c(\v_c^n)$ is the momentum source term containing contributions from all operators in \cref{eq:momentum-transport} from time $t^n$.

\Cref{eq:voldiscr,eq:momdiscrcoeff,eq:alphadiscr} cannot be solved simultaneously at the new time step $t^{n+1}$, because \cref{eq:voldiscr,eq:momdiscrcoeff} are linear algebraic equations resulting from an implicit UFVM discretization, and \cref{eq:alphadiscr} is solved geometrically. To ensure that equations \cref{eq:voldiscr,eq:momdiscrcoeff,eq:alphadiscr} are satisfied at $t^{n+1}$, we select a segregated solution algorithm developed by \OF as basis, i.e. PIMPLE, which combines SIMPLE (Semi-Implicit Method for Pressure-Linked Equations \citep{patankar1980numerical,patankar1972calculation}) and PISO (Pressure Implicit with Splitting of Operators \citep{issa1986solution}), to solve these equations, sequentially iterating until all equations are satisfied. 

We provide a simplified description of the PIMPLE algorithm, focusing on the interplay between the curvature approximation accuracy, surface tension force approximation and mesh non-orthogonality as the primary sources of numerical instabilities. 

Dividing semi-discrete momentum equation \cref{eq:momdiscrcoeff} by the coefficients $a_c$, and applying Rhie-Chow interpolation \citep{Rhie1983} gives
    \begin{equation}
         \v_f^{i} = - \left(\frac{1}{a_c}\right)_f (\nabla p)_f^{i} - \left(\frac{1}{a_c}\right)_f ((\g \cdot \x)\nabla \rho)_f^{o}
        - \left(\frac{1}{a_c}\right)_f\sigma_f (\kappa\nabla \alpha)_f^{o} + \left(\frac{1}{a_c}\right)_f(\mathbf{H}(F_f^o,\v^{i-1}))_f,
        \label{eq:facemomdiscrcoeff}
    \end{equation}
where $o$ denotes the outer iteration in which the momentum equation is discretized (and solved), and $i$ denotes the inner iterations used for solving the pressure Poisson equation derived below, and $\mathbf{H}(F_f^o,\v^{i-1}):= -\sum_{n\in N} a_n \v_n^{i-1} + \mathbf{S}_c(\v^n)$, with $a_n$ containing, among other terms, $F_f^o$, the volumetric flux resulting from the integration of \cref{eq:fluxdef} with second-order accurate face-centered quadrature $F_f^{o}=\v_f^{o}\cdot\Sf$.

Applying \cref{eq:voldiscr} to \cref{eq:facemomdiscrcoeff}, leads to the discrete Poisson equation for the pressure
    \begin{equation}
    \begin{aligned}
    \sum_{f \in F_c}
        \left( \frac{1}{a_c} \right)_f (\nabla p)^{i}_f \cdot \mathbf{S}_f
        & =
        - \sum_{f \in F_c} 
            \left(\frac{1}{a_c}\right)_f (\g \cdot \x)_f 
            (\nabla \rho)_f^{o} \cdot \mathbf{S}_f \\ 
        &    - \sum_{f \in F_c} 
            \left(\frac{1}{a_c} \right)_f 
            \sigma_f \kappa_f^{o} 
            (\nabla \alpha)_f^{o} \cdot \mathbf{S}_f \\
        &  + \sum_{f \in F_c}
            \left(\frac{1}{a_c}\right)_f 
            \mathbf{H}(F_f^{o}, \v^{i-1})_f\cdot \mathbf{S}_f. 
    \end{aligned}
    \label{eq:pressurefull}
    \end{equation}
The volume fraction, velocity, and pressure are coupled in \cref{eq:alphadiscrfromchi,eq:facemomdiscrcoeff,eq:pressurefull}, respectively, with the volume fraction $\alpha_c^o$ available from the solution of \cref{eq:alphadiscr} using the plicRDF-isoAdvector method, or any other method that approximates $\chi(\x,t)$ as a discrete phase-indicator $\alpha$ in the form of volume fractions, or, similarly, a marker field \citep{Liu2023}. 

\subsection{Force Balance} 

Regarding $(\nabla p)^{i}_f$, $(\g \cdot \x)_f(\nabla \rho)_f^{o}$ and $\sigma_f \kappa_f^{o}(\nabla \alpha)_f^{o}$: it is widely known that all face-centered gradients must be discretized with the same scheme to ensure a force-balanced discretization, that can be shown for the balanced CSF model \citep{Popinet2018}. 

To show that the Unstructured Finite Volume discretization is also balanced, we consider first a sphere suspended in zero-gravity with a constant mean curvature $\kappa$ in a steady state, which reduces \cref{eq:pressurefull} to 
\begin{equation}
    \sum_{f \in F_c}
        \left( \frac{1}{a_c} \right)_f^o (\nabla p)^{i}_f \cdot \mathbf{S}_f =  -\sum_{f \in F_c} \left(\frac{1}{a_c} \right)_f^o \sigma_f\kappa(\nabla \alpha)_f^{o} \cdot \mathbf{S}_f
        \label{eq:balanceSTpressure}
\end{equation}
i.e.,
\begin{equation}
    (\nabla p)^{i}_f + \sigma_f\kappa(\nabla \alpha)_f^{o}=0,
\end{equation}
and if the same discretization scheme $(\nabla.)_f$ is used, then following holds for a sphere
\begin{equation}
    \nabla(p^i + \sigma\kappa\alpha^o)=0,
\end{equation}
finally leading to 
\begin{equation}
    p^i=p^{C,i} - \sigma\kappa^{o}\alpha^o,
    \label{eq:youngLaplace}
\end{equation}
where $p^{C,i}$ is any constant pressure. \Cref{eq:youngLaplace} recovers the exact Young-Laplace pressure-jump across the interface $\Sigma(t)$, equivalently to structured discretizations \citep{Popinet2018}. If we approximate curvature as $\kappa^a$ with some approximation error $\epsilon_{\kappa}$, i.e., $\kappa^a=\kappa + \epsilon_{\kappa}$ and balance the forces using 
\begin{equation}
    (\nabla p)^{i}_f + \sigma_f\kappa_f^{a}(\nabla \alpha)_f=0,
\end{equation}
while considering only orthogonal gradient discretization, this results in 
\begin{equation}
\begin{aligned}
    \nabla(p + \sigma\kappa\alpha) & = -\sigma\epsilon_{\kappa}(\nabla\alpha)_f, \\ 
    \frac{p'^{,i}_{N_{f}} - p'^{,i}_{O_{f}}}{|\mathbf{d}_f|} &= -\frac{\sigma\epsilon_{\kappa}(\alpha^o_{N_{f}} - \alpha^o_{O_{f}})}{|\mathbf{d}_f|},  \\
    p'^{,i}_{N_{f}} &= p'^{,i}_{O_{f}} - \sigma\epsilon_{\kappa}(\alpha^o_{N_{f}} - \alpha^o_{O_{f}}), 
\end{aligned}
\label{eq:balancedapproxpressure}
\end{equation}
where $p' = p + \sigma\kappa\alpha$, and $O_f$, $N_f$ are the indices of cell centers from cells adjacent to the face $\mathbf{S}_f$. \Cref{eq:balancedapproxpressure} shows that modification of the Young-Laplace pressure, ensured for constant $\kappa$ by \cref{eq:youngLaplace} in the order-of-accuracy of $\nabla(.)_f$, is linearly proportional to the curvature approximation error $\epsilon_{\kappa}$, making curvature approximation crucial.

Curvature approximation in the context of numerical methods for two-phase flows is a long-standing challenge, which we do not address here. Instead, here we focus on ensuring that the unstructured FVM remains force-balanced on non-orthogonal meshes that are ubiquitous to industrially relevant two-phase flow problems that have geometrically complex flow domains.

\subsection{Force Balance on Non-Orthogonal Meshes} \label{sec:forceBalance}

\begin{figure}
    \def\svgwidth{.75\textwidth}
    {\footnotesize
\begingroup%
  \makeatletter%
  \providecommand\color[2][]{%
    \errmessage{(Inkscape) Color is used for the text in Inkscape, but the package 'color.sty' is not loaded}%
    \renewcommand\color[2][]{}%
  }%
  \providecommand\transparent[1]{%
    \errmessage{(Inkscape) Transparency is used (non-zero) for the text in Inkscape, but the package 'transparent.sty' is not loaded}%
    \renewcommand\transparent[1]{}%
  }%
  \providecommand\rotatebox[2]{#2}%
  \newcommand*\fsize{\dimexpr\f@size pt\relax}%
  \newcommand*\lineheight[1]{\fontsize{\fsize}{#1\fsize}\selectfont}%
  \ifx\svgwidth\undefined%
    \setlength{\unitlength}{705.29302658bp}%
    \ifx\svgscale\undefined%
      \relax%
    \else%
      \setlength{\unitlength}{\unitlength * \real{\svgscale}}%
    \fi%
  \else%
    \setlength{\unitlength}{\svgwidth}%
  \fi%
  \global\let\svgwidth\undefined%
  \global\let\svgscale\undefined%
  \makeatother%
  \begin{picture}(1,0.39738101)%
    \lineheight{1}%
    \setlength\tabcolsep{0pt}%
    \put(0,0){\includegraphics[width=\unitlength,page=1]{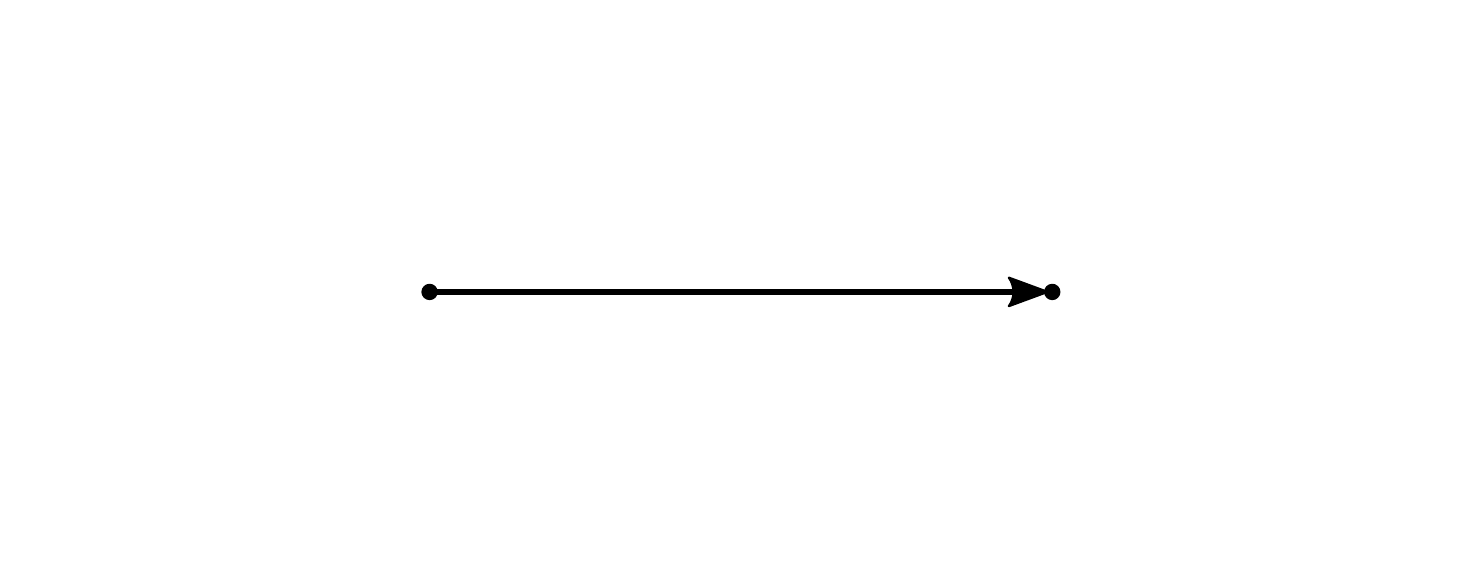}}%
    \put(0.28779595,0.17466426){\color[rgb]{0,0,0}\makebox(0,0)[lt]{\lineheight{1.25}\smash{\begin{tabular}[t]{l}$\mathbf{x}_{O_f}$\end{tabular}}}}%
    \put(0.71079053,0.17498359){\color[rgb]{0,0,0}\makebox(0,0)[lt]{\lineheight{1.25}\smash{\begin{tabular}[t]{l}$\mathbf{x}_{N_f}$\end{tabular}}}}%
    \put(0,0){\includegraphics[width=\unitlength,page=2]{def_nonOrthog.pdf}}%
    \put(0.45499761,0.17398642){\color[rgb]{0,0,0}\makebox(0,0)[lt]{\lineheight{1.25}\smash{\begin{tabular}[t]{l}$f$\end{tabular}}}}%
    \put(0.57246054,0.2731334){\color[rgb]{0,0,0}\makebox(0,0)[lt]{\lineheight{1.25}\smash{\begin{tabular}[t]{l}$\mathbf{S}_f$\end{tabular}}}}%
    \put(0,0){\includegraphics[width=\unitlength,page=3]{def_nonOrthog.pdf}}%
    \put(0.41903002,0.20506358){\color[rgb]{0,0,0}\makebox(0,0)[lt]{\lineheight{1.25}\smash{\begin{tabular}[t]{l}$\df$\end{tabular}}}}%
    \put(0.5598492,0.20978646){\color[rgb]{0,0,0}\makebox(0,0)[lt]{\lineheight{1.25}\smash{\begin{tabular}[t]{l}$\theta_f$\end{tabular}}}}%
    \put(0,0){\includegraphics[width=\unitlength,page=4]{def_nonOrthog.pdf}}%
  \end{picture}%
\endgroup%

    }
    \caption{Representation of non-orthogonality: $\O$, $\N$ are the centroids of two adjacent cells $O$, $N$; $\df$ is the vector connecting $\O$ and $\N$. }
    \label{fig:def_nonOrthog}
\end{figure}
\begin{figure}
    \def\svgwidth{\textwidth}
    {\footnotesize
     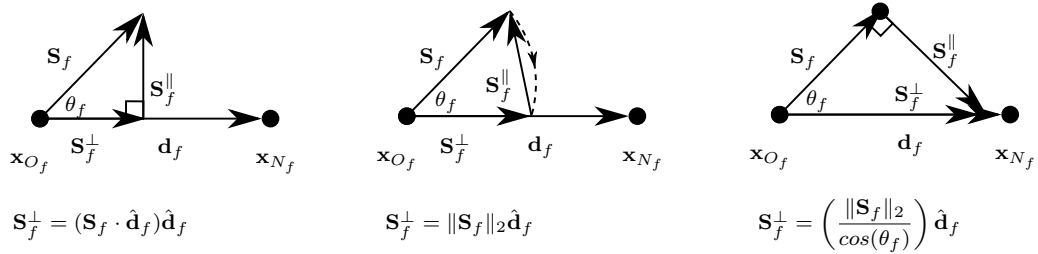
    }
    \caption{Three common non-orthogonality correction approaches: minimum correction (left), orthogonal correction (middle), over-relaxed correction(right). Vector $\dfhat$ is the unit vector of $\df$, i.e., $\dfhat := \frac{\df}{|\df|}$.}
    \label{fig:non-orthogonality-approaches}
\end{figure}

The unstructured Finite Volume Method must retain second-order accuracy and force balance also on non-orthogonal meshes: where the face area-normal vector $\Sf$ is not collinear with the vector connecting the centroids of cells, that share $\Sf$, i.e., $\mathbf{d}_f:=\x_{Nf} 
- \x_{Of}$, as depicted in \cref{fig:def_nonOrthog}. A widely used approach for non-orthogonality correction uses the principle of error superposition, i.e. splitting the total gradient flux into an explicit non-orthogonal $\|$ and implicit orthogonal $\perp$ contribution. Several non-orthogonal corrections \citep{Jasak1996,Muzaferija1997,Demirdvzic2015,Darwish2017},   
\begin{equation}
    (\nabla \phi)_f\cdot\Sf = (\nabla \phi)_f^{\perp}\cdot\Sf^{\perp} + (\nabla \phi)_f^{\|}\cdot\Sf^{\|},
    \label{eq:composimplexpl}
\end{equation}
for a property $\phi$, as shown in \cref{fig:non-orthogonality-approaches}.  A runtime-configurable simulation software such as \OF \citep{OFprimer} enables very straightforward consistent gradient scheme selection at the start of a simulation.

Following the principle of error superposition, if $(\nabla p)_f^{\|}\cdot\Sf^{\|}$ balances out $\sigma_f\kappa_f(\nabla \alpha)_f^{\|} \cdot \mathbf{S}_f^{\|}$ on the right hand side of \cref{eq:balanceSTpressure}, after applying \cref{eq:composimplexpl} in \cref{eq:balanceSTpressure} to $p$ and $\alpha$, the non-orthogonality correction will be force-balanced.  The logic following the force-balanced orthogonal gradient is that the same scheme used to discretize $(\nabla p)_f^{\|}\cdot\Sf^{\|}$ and $\sigma_f\kappa_f(\nabla \alpha)_f^{\|} \cdot \mathbf{S}_f^{\|}$ will ensure force-balance. 

However, ensuring that all terms are corrected in the same way is insufficient for force balance in the context of two-phase flow simulations. Since the non-orthogonality correction is explicit, $\sigma_f\kappa_f(\nabla \alpha)_f^{\|} \cdot \mathbf{S}_f^{\|}$ lags behind internal iterations in \cref{eq:pressurefull}. Focusing only on the force balance between the pressure gradient and the surface tension force in \cref{eq:pressurefull}, an explicit non-orthogonality correction becomes
\begin{equation}
    \begin{aligned}
    \sum_{f \in F_c}
        \left( \frac{1}{a_c} \right)_f^o (\nabla p)^{k,\perp}_f \cdot \mathbf{S}_f^{\perp}
        & =
           -\sum_{f \in F_c} 
            \left(\frac{1}{a_c} \right)_f^o 
            \sigma_f \kappa_f^{o} 
            (\nabla \alpha)_f^{o, \perp} \cdot \mathbf{S}_f^{\perp} \\
        & - \sum_{f \in F_c} 
            \left(\frac{1}{a_c} \right)_f^o 
            \sigma_f \kappa_f^{o} 
            (\nabla \alpha)_f^{o, \|} \cdot \mathbf{S}_f^{\|}\\
        & - \sum_{f \in F_c}
        \left( \frac{1}{a_c} \right)_f^o (\nabla p)^{k-1,\|}_f \cdot \mathbf{S}_f^{\|}.
    \end{aligned}
    \label{eq:orthogcompo}
\end{equation}
In \cref{eq:orthogcompo}, the explicit non-orthogonality correction indexed with $k-1$ lags behind the implicit orthogonal pressure gradient indexed with $k$ in balancing orthogonal and non-orthogonal $(\nabla\alpha)_f\cdot \Sf$ from the outer iteration $o$. This lag, if not resolved by absolutely ensuring a sufficient number of $k$ iterations are applied, causes oscillations in the pressure and, in turn, large parasitic velocities on non-orthogonal meshes, which perturb the fluid interface (volume fractions), reflecting the errors further into the curvature approximation, which closes the loop by forwarding the errors again into the pressure through \cref{eq:pressurefull}. 

We propose a deterministic stopping criterion for the $k$ non-orthogonality iteration, which ensures force-balance for non-orthogonal meshes that avoids introducing a problem-specific number of corrections, and significantly modifying the UFVM, or the solution algorithm. The solution is straightforward and becomes clear if the linear system given by \cref{eq:pressurefull}, extended with the non-orthogonality correction from \cref{eq:orthogcompo}, is written in matrix form as 
\begin{equation}
    \mathbf{L}^o \mathbf{p}^{k} = \mathbf{b}^o + \mathbf{s}_{no}^{k-1}
\end{equation}
where $\mathbf{p}$ (pressure), $\mathbf{b}$ (source term), $\mathbf{s}_{no}$ (non-orthogonality source) are cell-centered fields of the size equal to the number of cells in the mesh ($|C|$), and
\begin{equation}
    \mathbf{s}_{no}^{k-1} = -\sum_{f \in F_c} 
            \left(\frac{1}{a_c} \right)_f^o 
            \sigma_f \kappa_f^{o} 
            (\nabla \alpha)_f^{o, \|} \cdot \mathbf{S}_f^{\|} 
            - \sum_{f \in F_c}
        \left( \frac{1}{a_c} \right)_f^o (\nabla p)^{k-1,\|}_f \cdot \mathbf{S}_f^{\|}
\end{equation}
is the force-imbalance between the pressure gradient and the surface-tension force resulting from the explicit non-orthogonal correction. The source term $\mathbf{s}_{no}$ should diminish with increasing $k$, independent of the chosen non-orthogonality correction, as long as the same correction is applied to all gradients, and the correction is, of course, convergent.  
A deterministic way to ensure that $\mathbf{s}_{no}^k$ diminishes, is by 
\begin{equation}
    |\mathbf{L}^o \mathbf{p}^{k} - \mathbf{b}^o - \mathbf{s}_{no}^{k-1}|_{\lambda} < \tau,
    \label{eq:balancePressureErrorNorm}
\end{equation}
i.e., requiring the non-orthogonal force-balance to reach the accuracy of the linear-solver tolerance $\tau_S$ using a linear-solver residual norm $\lambda_S$, for the pressure Poisson equation. \Cref{eq:balancePressureErrorNorm} works because of the way modern linear solver algorithms are implemented. Given that, in Computational Fluid Dynamics, we are solving PDEs whose solutions evolve either over iterations for steady state problems, or time steps for transient problems, in each subsequent iteration, the linear solver will first compute the initial residual using the solution from the previous iteration. If the chosen non-orthogonality correction is convergent, with increasing $k$, it will achieve force-balance. At the point of force-balance, the linear solver will use the current $\mathbf{p}^k$ to compute the residual of \cref{eq:pressurefull}, which will fall under the tolerance $\tau$ given the solver norm $\lambda$. This defines a straightforward and entirely deterministic termination criterion for a force-balanced explicit non-orthogonal correction, and does so, without the need to quantify the tolerance for the non-orthogonality error. In addition to being deterministic, the residual-based non-orthogonality correction in \cref{eq:balancePressureErrorNorm} works in the same way for any linear solver and any tolerance. We also claim that \cref{eq:balancePressureErrorNorm} will incur a minimal number of corrections - the force-balance cannot be more accurate than satisfying \cref{eq:pressurefull} using the chosen linear solver norm and tolerance. If the non-orthogonality error is small,  $\mathbf{s}_{no}^{k-1}$ will diminish quickly, if it is large, an increase of iterations is justified and confirmed by our results section. 

Satisfying \cref{eq:balancePressureErrorNorm} is very straightforward in a numerical source code: it is equivalent to expecting that the linear solver exits the initial iteration since this means that the initial $p^k$ used to compute the initial residual, already satisfies \cref{eq:balancePressureErrorNorm}.

We designate the proposed non-orthogonality correction as Residual-based Non-Orthogonality Correction (\emph{ResNonOrthCorr}) and outline the required straightforward modification of the segregated solution algorithm to incorporate ResNonOrthCorr in \cref{alg:ResNonOrthCorr}.

In \cref{alg:ResNonOrthCorr} we add the option to stop if the number of iterations exceeds $N_{max}$ as means of avoiding exceedingly large number of corrections that would lead to computationally intractable simulations, in cases with unavoidable extremely large non-orthogonality in industrial applications. 

When adding the gravity force to balanced forces, equivalently to the CSF surface tension force, we apply the same principle of error-superposition and extend  $\mathbf{s}_{non}^k$ to include the explicit non-orthogonal gravity force contribution
\begin{equation}
    \begin{aligned}
    \mathbf{s}_{no}^{k-1} & = -\sum_{f \in F_c} 
            \left(\frac{1}{a_c} \right)_f^o 
           \left[ \sigma_f \kappa_f^{o} 
        + 
        (\rho^- - \rho^+)
            (\mathbf{g}\cdot \mathbf{x})_f\right] 
            (\nabla \alpha)_f^{o, \|} \cdot \mathbf{S}_f^{\|} \\
        & -  \sum_{f \in F_c}
        \left( \frac{1}{a_c} \right)_f^o (\nabla p)^{k-1,\|}_f \cdot \mathbf{S}_f^{\|}.
    \end{aligned}
        \label{eq:nonorthsourcefinal}
\end{equation}
It is important to note at this point that, from \cref{eq:rhoindicator},
\begin{equation}
    \nabla \rho = (\rho^ - - \rho^+) \nabla \chi \approx (\rho^- - \rho^+) \nabla \alpha.
\end{equation}
With sufficient information about the UFVM discretization, segregated solution algorithms and non-orthogonal force balance, we proceed to verify ResNonOrthCorr.

\section{Comparison with state-of-the-art methods}
\label{sec:state-of-the-art}

We first compare ResNonOrthCorr with the widely used heuristic fixed-number of non-orthogonality corrections (\emph{FixNonOrthCorr}) in the PIMPLE \cref{alg:PIMPLE} algorithm in OpenFOAM \citep{OF_V2306} and similar iterative segregated solution algorithms.

The PIMPLE method updates $\alpha_c,\v_c$ and  $p_c$ as outlined in \cref{alg:PIMPLE}. In each outer loop, the geometric information of the interface and physical properties are updated. In the inner loop, source terms of \cref{eq:pressurefull} are calculated and \cref{eq:pressurefull} is solved.  

\begin{center}
\begin{algorithm}[H]
    \centering
    \caption{FixNonOrthCorr in the PIMPLE solution algorithm.}
    \label{alg:PIMPLE}
    {\small
    \begin{algorithmic}[1]
        \While{ $t \le t_{end}$ }
            \State $t^{n+1} = t^{n} + \Delta t$ 
            \For{$o = 1$; $o \le N_{outer}$; $++o$} 
                \State Reconstruct the fluid interface, i.e., $\tilde{\chi}(\x,t^{o-1})\approx \chi(\x,t^{o-1})$ from $\alpha_c^{o-1}$.
                \State Solve \cref{eq:alphadiscrfromchi} for $\alpha_c^{o}$ using $\tilde{\chi}(\x,t^{o-1}), F_f^{o-1}$, giving $\rho_{f,c}^{o},\mu_f^{o}$ from \cref{eq:rhoindicator,eq:nuindicator}, respectively. 
                \State Discretize the momentum \cref{eq:momdiscrcoeff} using $\alpha_c^o, \rho_c^o, \rho_f^o, \mu_f^o$ compute the $\H(F_f^{o-1}, \v^{o-1})$ operator. 
                \For{$i = 1$; $i \le N_{inner}$; $++i$} 
                    \For{$k = 1$; $k \le N_{non}$; $++k$} \Comment Non-orthogonality correction.
                        \State Solve the pressure equation \cref{eq:pressurefull} for $p_k^i$ with $\H(F_f^{o},\v^{i-1})$ and $\alpha_c^o, \rho_c^o$.
                    \EndFor
                    \State Update $F_f^i, \v_c^i$ with $\H(F_f^{o},\v^{i-1})$ and $p_c^{Non} using $
                \EndFor
            \EndFor
        \EndWhile
    \end{algorithmic}    
    }
    \end{algorithm}
\end{center}

\begin{center}
\begin{algorithm}[H]
    \centering
    \caption{ResNonOrthCorr in the PIMPLE algorithm.}
    \label{alg:ResNonOrthCorr}
    {\small
    \begin{algorithmic}[1]
        \While{ $t \le t_{end}$ }
            \State $t^{n+1} = t^{n} + \Delta t$ 
            \For{$o = 0$; $o < N_{outer}$; $++o$} 
               \State Reconstruct the fluid interface, i.e., $\tilde{\chi}(\x,t^{o-1})\approx \chi(\x,t^{o-1})$ from $\alpha_c^{o-1}$.
                \State Solve \cref{eq:alphadiscrfromchi} for $\alpha_c^{o}$ using $\tilde{\chi}(\x,t^{o-1}), F_f^{o-1}$, giving $\rho_{f,c}^{o},\mu_f^{o}$ from \cref{eq:rhoindicator,eq:nuindicator}, respectively. 
                \State Discretize the momentum \cref{eq:momdiscrcoeff} using $\alpha_c^o, \rho_c^o, \rho_f^o, \mu_f^o$ compute the $\H(F_f^{o-1}, \v^{o-1})$ operator.  
                \State $k = 0$
                \For{$i = 0$; $i < N_{inner}$; $++i$} 
                    \While{$|L^0p^{k+1} - b^o - \mathbf{s}_{no}^k|_{\lambda} > \tau$ or $k > N_{max}$} 
                        \State $k = k + 1$
                        \State Solve the pressure equation \cref{eq:pressurefull} for $p_k^i$ with $\H(F_f^{i-1},\v^{i-1})$ and $\alpha_c^o, \rho_c^o$.
                    \EndWhile
                    \State Update $F_f^o, \v_c^o$ with $F_f^i, \v_c^i$ from $p_k^i$ using \cref{eq:facemomdiscrcoeff}.
                \EndFor
            \EndFor
        \EndWhile
    \end{algorithmic}    
    }
    \end{algorithm}
\end{center}
Correcting face-centered gradients for non-orthogonality introduces the non-orthogonality correction loop $k$ in \cref{alg:PIMPLE}. Generally, $N_{outer}$, $N_{inner}$ and $N_{non}$, are set by the user of \cref{alg:PIMPLE}, making $N_{non}$ a problem-specific "free" parameter. The number $N_{non}$ in \cref{alg:PIMPLE} requires adjustment by trial and error; we show in the results section that this is insufficient for achieving force-balance. It is widely known that the pressure Poisson \cref{eq:pressurefull} is the computational bottleneck in CFD. In segregated solution algorithms, the computational cost is directly proportional to the number of times  \cref{eq:pressurefull} is solved, namely, $N_{outer} \times N_{inner} \times N_{non}$. The ResNonOrthCorr completely removes $N_{non}$ as a "free" parameter from the solution \cref{alg:PIMPLE} and replaces the stopping condition in the innermost loop in \cref{alg:ResNonOrthCorr}.

A recent algorithm proposed by \citep{Huang2023} seeks to balance $(\nabla p)_f$ with $(\nabla \alpha)_f$ on a non-orthogonal mesh by employing the same scheme to discretize pressure and forces gradients. This approach combines explicit correction terms for both discretized pressure and forces gradients into a "revised" pressure gradient. The reconstruction of this new pressure gradient at the cell center is accomplished using, what the authors call, the Time-evolution Converting (TEC) operator, instead of the conventional Green Gauss and least-square methods. The TEC operator, proposed in \citep{Xie2017unstructured}, is actually a variant of the least-square method that minimizes the sum of squares of the error between the computed fluxes from the cell center gradient and the directly estimated fluxes at the cell faces. This has been in widespread application in \OF for reconstructing cell center values from known fluxes, as discussed by \citet{Aguerre2018,Tolle2020} and revisited by \citet{Assam2021novel}. 

Prior works, such as those by \citep{Patel2017,Xie2017unstructured,Manik2018,Xie2020}, have also acknowledged the impact of consistent discretization between pressure gradient flux and force gradient fluxes in preserving force balance on non-orthogonal meshes. Consequently, they utilized identical discretized formulations for all these gradient fluxes. However, the correction for non-orthogonality in these methods occurs only once per internal iteration when solving the discretized Poisson equation, i.e., during the assembly of the coefficients matrix and sources of the discretized Poisson equation. 

The discretization of gradient flux (\cref{eq:composimplexpl}) has been extensively discussed, not only for pressure and force gradients but also for diffusive fluxes in various fields such as velocity, temperature, and electric fields, among others. Over the past decades, numerous methods have been developed to address this discretization challenge \citep{Demirdvzic2015}. \citet{Muzaferija1994,Muzaferija1997} initiated an approach to approximate the gradient flux by interpolating the gradients at two adjacent cell centers to the intersected face and multiplying the interpolated face gradient by the face area vector. However, they found that this simple gradient interpolation method could introduce unphysical oscillations in the solution on a collocated mesh. To mitigate this issue, recoupling terms, inspired by oscillatory pressure handling methods proposed by \citet{Rhie1983}, were added to the interpolated face gradient \citep{Muzaferija1994,Muzaferija1997}. \citet{Demirdvzic1995numerical} made slight adaptations to these added terms \citep{Muzaferija1994,Muzaferija1997} to ensure their vanishing when the solution converges. \citet{Nishikawa2010beyond,Nishikawa2011robust} proposed a general principle for discretizing the diffusion term based on a first-order hyperbolic system. In the context of the finite volume method, the diffusion discretization scheme involves the addition of an arithmetic average of neighboring cell center gradients and a jump term containing a free parameter. This parameter is used to dampen high-frequency errors caused by the arithmetic average. In addition to these pure numerical derivations for diffusion discretization, some works (\citep{Jasak1996,Darwish2017,Ferziger2020}) chose a geometric perspective. \citet{Jasak1996} decomposed the face area vector $\mathbf{S}_f$ into the direction of the line connecting two neighboring cell centers, denoted as $\mathbf{d}_f$ in \cref{fig:non-orthogonality-approaches}, and the vector $\mathbf{S}_f - \mathbf{d}_f$.  The central differencing formula was then used to implicitly approximate the product of the face gradient and the $\mathbf{d}_f$ component, while face interpolation of the center gradient was used to explicitly approximate the product of the face gradient and the $\mathbf{S}_f - \mathbf{d}_f$ component. The author also compared these three correction methods in \cref{fig:non-orthogonality-approaches} and showed the over-relaxation correction has the best performance. \citet{Ferziger2020} extrapolated two neighboring cell center values to two auxiliary nodes, which are located on the line in the face normal direction and passing through the face center. The extrapolated values and positions of these two nodes are used to construct a central difference to approximate the face gradient. Similarly, \citet{Darwish2017} introduced a scheme utilizing two auxiliary nodes, although these two nodes are not placed on the normal line through face center. Instead, they are constrained by constructing two opposite directional gradient fluxes with the corresponding near cell centers. The final gradient flux is calculated by averaging the two fluxes. \citet{Darwish2017} indicated that the gradient flux could be fully implicit by carefully selecting the averaging parameters concerning the node values and positions. For this fully implicit scheme, each internal iterative step in solving the discretized equation contained one non-orthogonal correction, leading to fewer iterations required to converge to a satisfactory tolerance compared with the semi-implicit schemes, where the correction is conducted only once during the assembly of the source term of the discretized equation. While numerous methods have been developed, review papers such as \citep{Wu2013similarity,Demirdvzic2015} reveal equivalences between the final formulations of the discretized diffusion flux in different works, e.g., \citep{Jasak1996} and \citep{Muzaferija1994,Muzaferija1997,Demirdvzic1995numerical}. A comprehensive comparison of various semi-implicit diffusion discretization schemes, considering both discretization and truncation errors, is presented in \citet{Jalali2014accuracy}. The authors \citep{Jalali2014accuracy} concluded that the most accurate approximation for diffusive fluxes is achieved by adding a solution jump term to the average of two adjacent cell gradients, as proposed in \citep{Nishikawa2010beyond,Nishikawa2011robust}.

All contemporary methods require a number of iterations for the non-orthogonality correction that is left as a free parameter. Our contribution to non-orthogonality correction using the gradient-flux decomposition, regardless of the chosen gradient scheme or the numerical method for tracking fluid interfaces, is a deterministically controlled number of iterations that ensures force-balance for two-phase flow simulations on non-orthogonal meshes, verified in the next section.

\section{Results}
\label{sec:validations-results}

Data archives including the algorithm implementation, input data, post-processing software and secondary data are publicly available \citep{nonOrthogCodes2023,nonOrthogData2023}. The proposed method is actively developed in a publicly available git repository \citep{nonOrthogRepository}.

\begin{figure}[!htb]
     \begin{subfigure}{.3\textwidth}
      \centering
      \includegraphics[width=\linewidth]{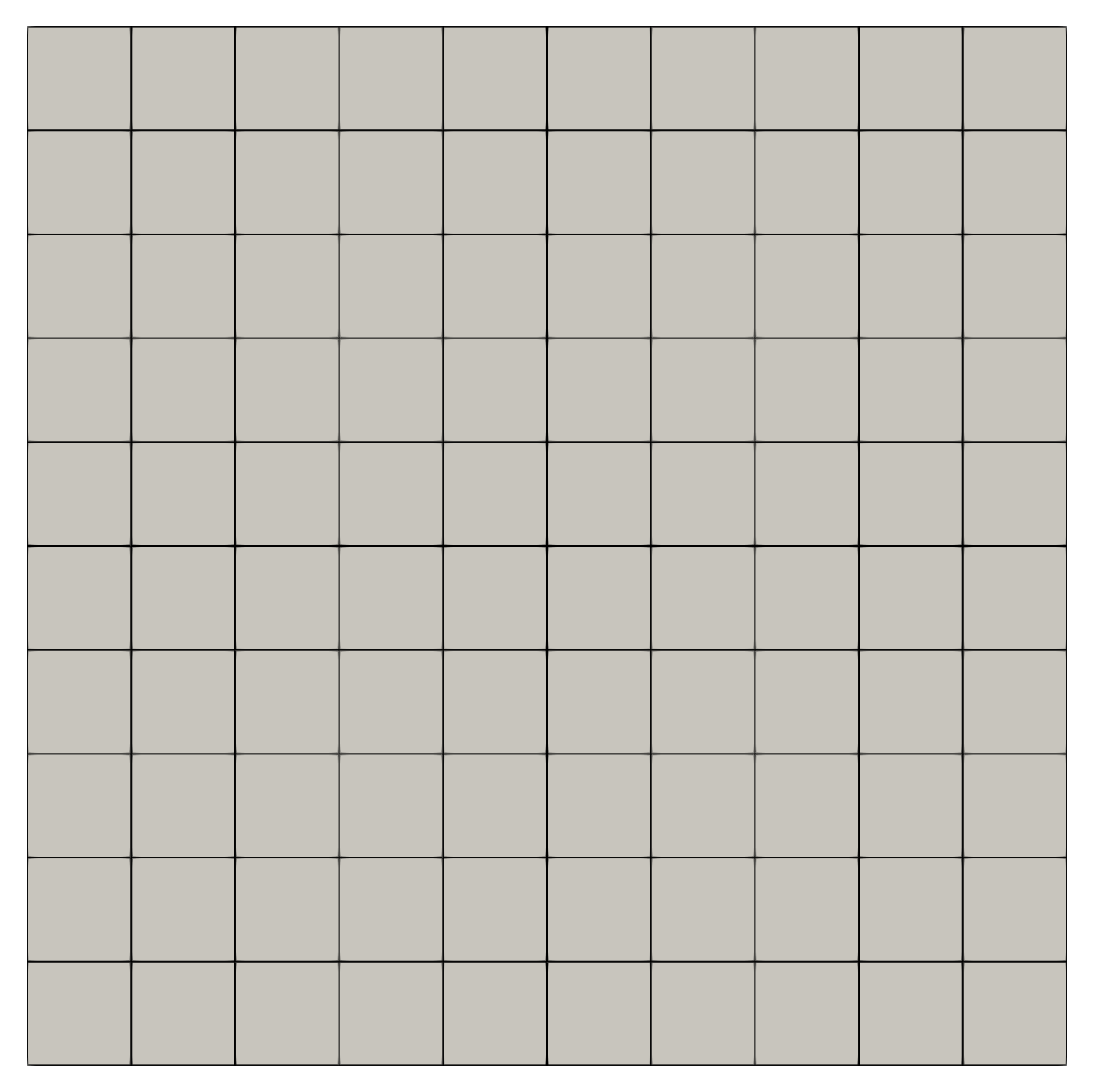}
      \caption{blockMesh}
     \end{subfigure}
     \begin{subfigure}{.3\textwidth}
      \centering
      \includegraphics[width=\linewidth]{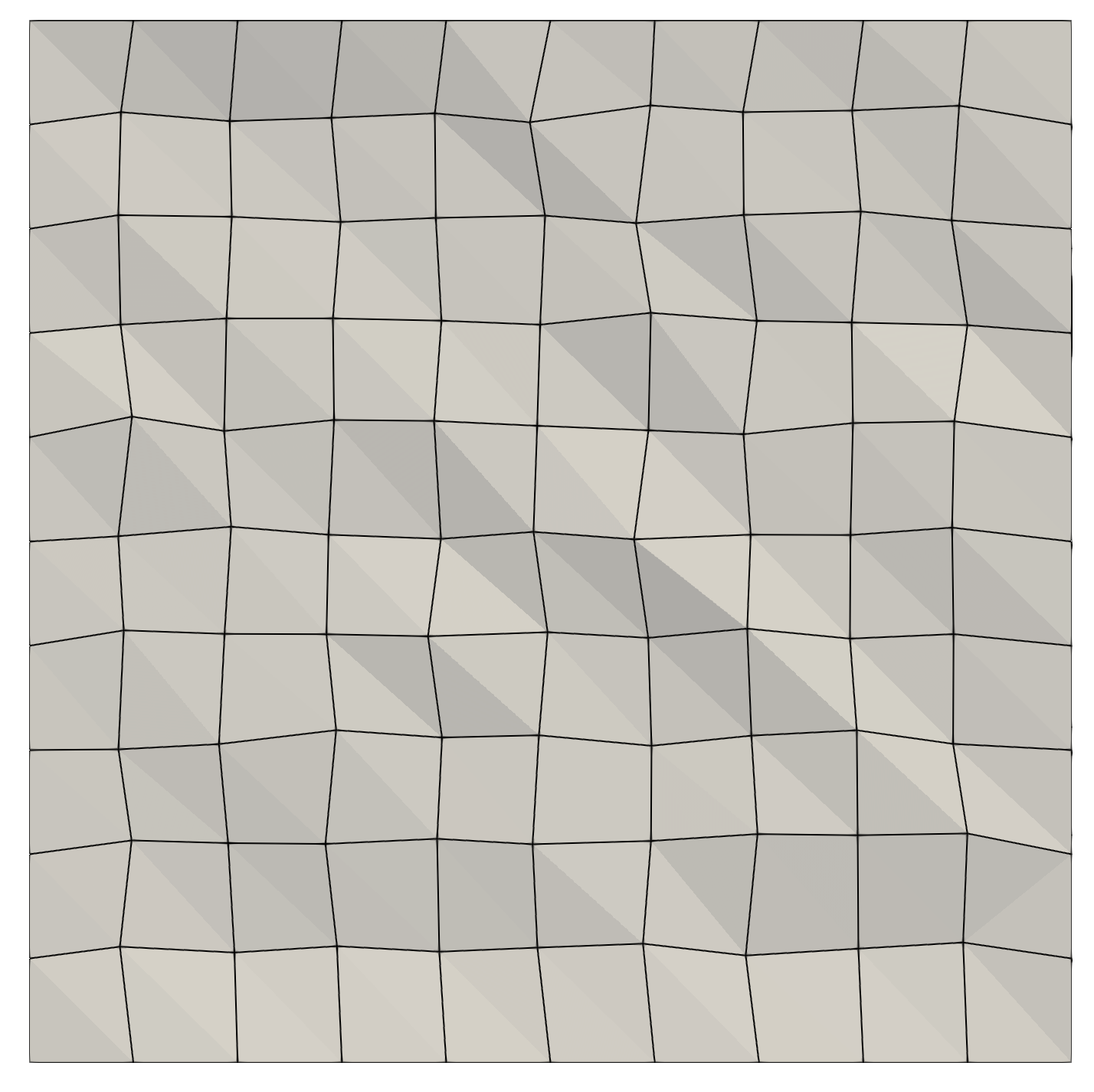}
      \caption{perturbMesh}
     \end{subfigure}
     \begin{subfigure}{.306\textwidth}
      \centering
      \includegraphics[width=\linewidth]{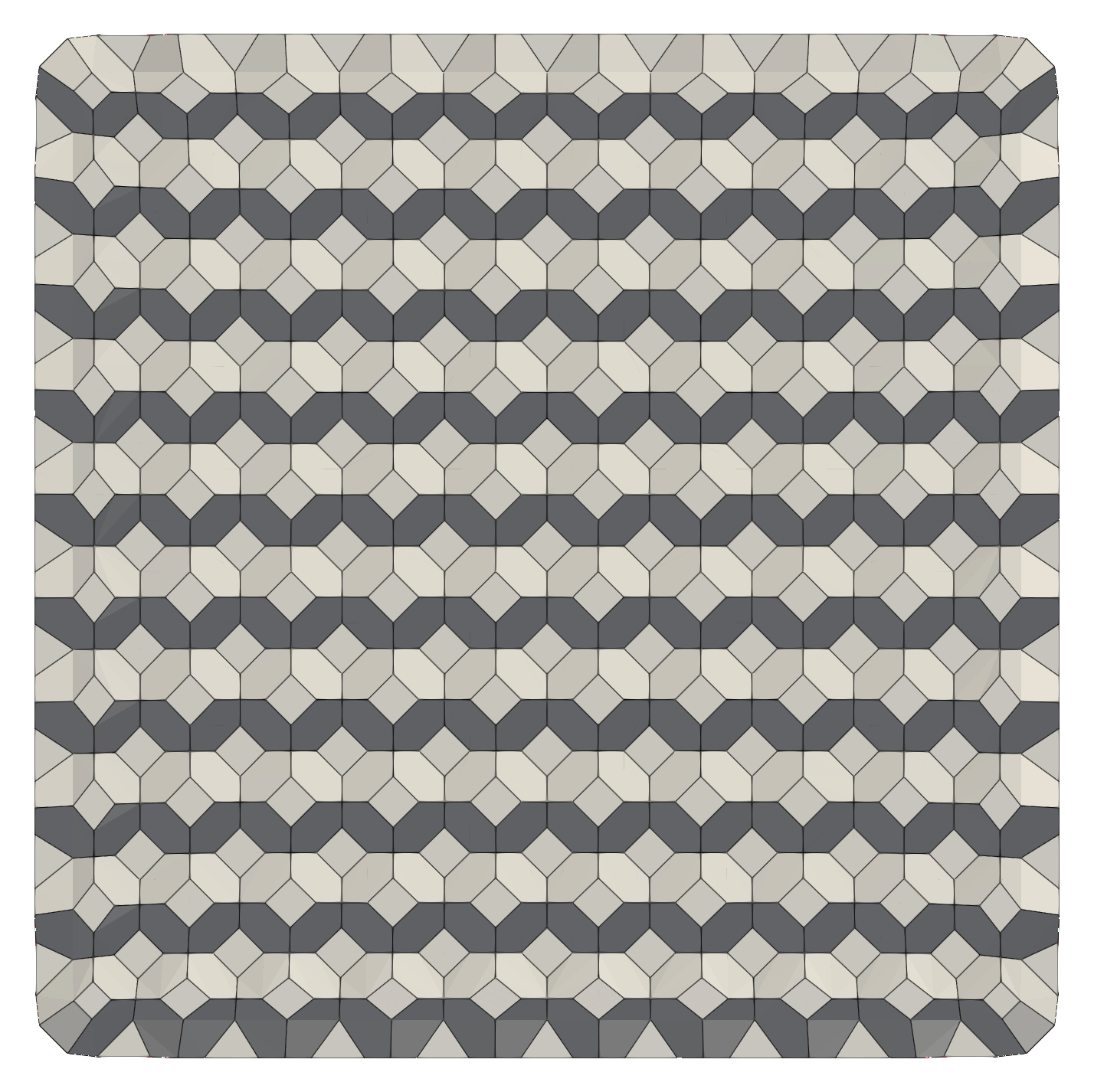}
      \caption{polyMesh}
     \end{subfigure}    
    \caption{A sliced cell layer from a cubic computational domain with three mesh types: equidistant mesh (\textit{blockMesh}), perturbed hexahedral mesh (\textit{perturbMesh}) and polyhedral mesh (\textit{polyMesh}).}
    \label{fig:mesh-types}
\end{figure}

In this section, we assess the efficiency and accuracy of the proposed force-balanced algorithm using two canonical verification tests, namely, the stationary droplet and the stationary water tank, used, respectively, to test the force-balanced surface tension force and gravity force discretization. These tests are conducted on unstructured meshes and the non-orthogonality corrections are either kept fixed (\emph{FixNonOrthCorr}) or are controlled by our residual-based algorithm (\emph{ResNonOrthCorr}). To systematically investigate the influence of different mesh types on the solutions, simulations are performed on equidistant unstructured mesh (\textit{blockMesh}), perturbed equidistant unstructured mesh, i.e. hexahedral mesh (\textit{perturbMesh}), and polyhedral mesh (\textit{polyMesh}), as illustrated in \cref{fig:mesh-types}. Three distinct mesh resolutions are investigated, with resolutions $\Delta x \in (L/30, L/60, L/90)$. Here, $L$ indicates the characteristic length of the cubic computational domain $\Omega$. These hydrodynamic test cases consider a water/air fluid pairing to emphasize a problematic two-phase flow scenario, the fluid properties are outlined in \cref{table:waterAirProperties}. 

Both tests share identical termination times and fixed time steps, specifically $t_{end}=\SI{0.01}{s}$ and $\Delta t= \SI[output-exponent-marker = e]{1e-4}{s}$. For the segregated solution algorithm \cref{alg:ResNonOrthCorr} we use, exemplary, $N_{outer}=4, N_{inner}= 1$. For FixNonOrthCorr, non-orthogonality is corrected $N_{non}$ times within an inner loop. In the case of ResNonOrthCorr, the non-orthogonality correction is performed until \cref{eq:balancePressureErrorNorm} is satisfied. The linear solver tolerance for \cref{eq:pressurefull} is uniformly set to $\num[output-exponent-marker = e]{1e-12}$ for both algorithms.

\begin{table}[]
\begin{tabular}{llllll}
\toprule
$20\ \unit{\degreeCelsius}$ & \begin{tabular}[c]{@{}l@{}}density $\rho$ \\ ($\unit{kg/m^3}$)\end{tabular} & \begin{tabular}[c]{@{}l@{}}kin. viscosity $\nu$ \\ ($\unit{m^2/s}$)\end{tabular} & \begin{tabular}[c]{@{}l@{}}dyn. viscosity $\mu$ \\ ($\unit{Pa.s}$)\end{tabular} & \begin{tabular}[c]{@{}l@{}}surface tension $\sigma$ \\ ($N/m$)\end{tabular} & \begin{tabular}[c]{@{}l@{}}gravity $\g$ \\ ($m/s^2$)\end{tabular} \\ \midrule
water\citep{Linstrom2005,VDI2010}                       & $998.2$                                                                     & $\num[output-exponent-marker = e]{1e-6}$                                                                           & $\num[output-exponent-marker = e]{9.982e-4}$                                                                                     & \multirow{2}{*}{$\num[output-exponent-marker = e]{72.74e-3}$\citep{Petrova2014revised}}                                                 & \multirow{2}{*}{$(0,0,-9.81)$}                                   \\
air\citep{VDI2010}                         & $1.19$                                                                      & $\num[output-exponent-marker = e]{1.53e-5}$                                                                        & $\num[output-exponent-marker = e]{1.8207e-5}$                                                                                    &                                                                             &                                                                  \\ \bottomrule
\end{tabular}
\caption{Physic properties of water/air pair.}
\label{table:waterAirProperties}
\end{table}

\subsection{Stationary droplet in equilibrium}
In accordance with the Young–Laplace law, the velocity of a spherical droplet in equilibrium, in the absence of gravity, is zero, i.e. $\v = 0$. This is attributed to the balance between the surface tension force and the pressure jump across the interface. This test involves a stationary water droplet, with microfluidic dimensions, in equilibrium with surrounding air. The water droplet of radius $R=1\ \unit{\mm}$ is placed at the centroid of the cubic domain $\Omega$, with domain dimensions $[0,0,0]\times[10,10,10]\ \unit{\mm}$. For a spherical droplet, the curvature is uniform throughout the interface, denoted as $\kappa_{\Sigma}=\frac{2}{R}= 2000\ \unit{m^{-1}}$. This curvature value is prescribed and held constant in simulations to avoid testing errors arising from curvature approximation and only verify the force-balanced non-orthogonality correction. Two error norms are evaluated to highlight the accuracy of the algorithm, namely
\begin{equation}
  \begin{aligned}
        L_{\infty}(|\v|) &= \text{max}(|\v_n - \v_e|) \\
        L_\infty(|\Delta p|) &= \frac{|\text{max}(\Delta p_n)-\Delta p_e|}{\Delta p_e}\\ 
        \text{max}(\Delta p_n) &= \text{max}(p)_n - \text{min}(p)_n
    \end{aligned}
    \label{eq:error_norms}
\end{equation}
where the subscript $n$ and $e$ denote the numerical and exact solutions correspondingly. The exact solutions are 
\begin{equation*}
  \begin{aligned}
        \v_e &= \mathbf{0}\ \unit{m/s}, \\
        \Delta p_e &= \sigma\kappa_{\Sigma} = 145.48\ \unit{Pa}.
    \end{aligned}
\end{equation*}
\begin{table}[]
\begin{adjustbox}{width=1\textwidth}
\small
\begingroup
\begin{tabular}{llllllllll}
\toprule
Mesh type                    & Resolution          &  \multicolumn{2}{l}{max. non-ortho.}                                      & \multicolumn{3}{l}{ResNonOrthCorr}                                                                                                                                                                                       & \multicolumn{3}{l}{FixNonOrthCorr($N_{non}=1$)}\\
                             & $\frac{\Delta x}{L}$ &\begin{tabular}[c]{@{}l@{}}global($\theta_f$) \\ ($\unit{\degree}$)\end{tabular} &\begin{tabular}[c]{@{}l@{}}local($\theta_f$) \\ ($\unit{\degree}$)\end{tabular}     & \begin{tabular}[c]{@{}l@{}}$L_{\infty}(|\v|)$\\ ($\unit{m/s}$)\end{tabular} & \begin{tabular}[c]{@{}l@{}}$L_\infty(|\Delta p|)$\\ ($\unit{Pa}$)\end{tabular} & \begin{tabular}[c]{@{}l@{}}CPU time\\ ($s$)\end{tabular} & \begin{tabular}[c]{@{}l@{}}$L_{\infty}(|\v|)$\\ ($\unit{m/s}$)\end{tabular} & \begin{tabular}[c]{@{}l@{}}$L_\infty(|\Delta p|)$\\ ($\unit{Pa}$)\end{tabular} & \begin{tabular}[c]{@{}l@{}}CPU time\\ ($s$)\end{tabular} \\ \midrule
blockMesh   & 1/30               &0  & 0                                                          & 2.4567e-11                                                                  & 1.4457e-14                                                              & \multicolumn{1}{l|}{432.32}                              & 9.1924e-11                                                                  & 8.2053e-15                                                              & 384.54                                                   \\
                             & 1/60           & 0      & 0                                                          & 1.7332e-11                                                                  & 1.7583e-14                                                              & \multicolumn{1}{l|}{3138.92}                             & 3.0472e-10                                                                  & 2.8914e-14                                                              & 3032.42                                                  \\
                             & 1/90           & 0      & 0                                                          & 7.9776e-12                                                                  & 2.3835e-14                                                              & \multicolumn{1}{l|}{8990.49}                             & 6.0617e-12                                                                  & 2.2076e-14                                                              & 8633.76                                                  \\ \midrule
perturbMesh & 1/30     &  12.79           & 10.09                                                      & 4.0199e-10                                                                  & 3.7432e-13                                                              & \multicolumn{1}{l|}{463.24}                              & 2.1683e-07                                                                  & 3.9894e-10                                                              & 560.91                                                   \\
                             & 1/60        & 14.45         & 10.87                                                      & 2.8729e-10                                                                  & 3.6006e-13                                                              & \multicolumn{1}{l|}{3294.92}                             & 1.5253e-06                                                                  & 4.063e-09                                                               & 5210.09                                                  \\ 
                             & 1/90         &13.54        & 11.00                                                      & 1.7006e-09                                                                  & 1.8767e-12                                                              & \multicolumn{1}{l|}{9432.08}                             & 1.8398e-06                                                                  & 5.8021e-09                                                              & 18328.48                                                 \\ \midrule
polyMesh    & 1/30          &30.81       & 1.48e-06                                                   & 3.9613e-11                                                                  & 4.6731e-13                                                              & \multicolumn{1}{l|}{1653.41}                             & 4.3341e-11                                                                  & 1.6684e-13                                                              & 1600.45                                                  \\
                             & 1/60        & 33.42        & 1.57e-06                                                   & 7.7984e-12                                                                  & 3.5556e-14                                                              & \multicolumn{1}{l|}{10256.68}                            & 9.0486e-12                                                                  & 1.2015e-13                                                              & 9630.83                                                  \\
                             & 1/90        & 30.81         & 1.71e-06                                                   & 7.3637e-12                                                                  & 1.6743e-13                                                              & \multicolumn{1}{l|}{38833.63}                            & 1.9106e-12                                                                  & 1.1917e-14                                                              & 37077.32                                                 \\ \bottomrule
\end{tabular}
\endgroup
\end{adjustbox}
\caption{Maximum global and near-interface non-orthogonality, the velocity and pressure jump errors, and CPU times for a stationary droplet in equilibrium at the end time $t_{end}=0.1s$}
\label{table:staticDroplet3D}
\end{table}
\Cref{table:staticDroplet3D} presents CPU time, the final velocity and pressure jump errors for both control algorithms on various meshes, along with information regarding non-orthogonality. The non-orthogonality is corrected once per inner loop for FixNonOrthCorr in \cref{table:staticDroplet3D}, denoted as $N_{non}=1$. The non-orthogonality at a cell face described in \cref{table:staticDroplet3D} is quantified by the angle of intersection of the line connecting two adjacent cell centers and the face normal, i.e., $\theta_f$ depicted in \cref{fig:def_nonOrthog}. Two maximum non-orthogonalities are documented in \cref{table:staticDroplet3D}. The global maximum non-orthogonality represents the largest $\theta_f$ among cell faces throughout the computational domain $\Omega$. Additionally, particular attention is given to the maximum non-orthogonality in the vicinity of the interface that actually causes force-imbalance resulting in parasitic currents. Specifically, we assess the maximum non-orthogonality over the faces in the interface cell-layer and two adjacent cell layers.     

When the pressure Poisson equation \cref{eq:pressurefull} is solved on orthogonal meshes (i.e, blockMesh or orthogonal polyMesh), where the non-orthogonality is zero or near-zero, both ResNonOrthCorr and FixNonOrthCorr perform similarly. Very minor disparities in two error norms and CPU time are observed, demonstrating a very low computational overhead of ResNonOrthCorr stopping criterion. On the orthogonal mesh, the explicit non-orthogonal contribution $(\cdot)^{\|}_f$ in \cref{eq:composimplexpl} is zero, implying that the non-orthogonal correction does not introduce new errors. The errors reported in \cref{table:staticDroplet3D} for cases employing blockMesh arise solely from the orthogonal contribution. Notably, the errors approach the prescribed error tolerance, reaffirming the force-balance property inherent in the consistent discretization of pressure and body force gradients at cell faces \citep{Francois2006}, for the unstructured Finite Volume method \citep{nonOrthogRepository}. 

The non-orthogonality of perturbMesh in \cref{table:staticDroplet3D} is uniformly distributed throughout the computational domain, with both global and local non-orthogonality ranging between $10\unit{\degree}$ and $15\unit{\degree}$ - still very small and generally acceptable magnitudes. Even for acceptably small non-orthogonality, strong differences arise between ResNonOrthCorr and FixNonOrthCorr. The final $L_{\infty}(|\v|)$ and $L_\infty(|\Delta p|)$ values computed with ResNonOrthCorr are three to four orders of magnitude smaller than those obtained with FixNonOrthCorr. Furthermore, the CPU times for all three cases with varying resolutions employing ResNonOrthCorr are markedly lower than those using FixNonOrthCorr, especially with a reduction in approximately half the CPU time for the two higher-resolution cases. The force-balance impacts computational efficiency: a force-balanced discretization will recover the accurate steady state, causing less work for the solution algorithm and the linear solver, compared to the lack of force balance. Not ensuring force balance incurs acceleration of the fluid, whose velocity changes in time (until possibly reaching steady state), and should be divergence-free for incompressible fluid, which means more work for the solution algorithm and the pressure Poisson equation. 

To thoroughly investigate the impact of the non-orthogonality correction number $N_{non}$ on the errors and CPU time for the FixNonOrthCorr algorithm, we conduct additional tests with two distinct groups of cases featuring identical perturbMesh setups as outlined in \cref{table:staticDroplet3D} but with larger values for the number of non-orthogonality corrections, i.e., $N_{non}=[2,10]$, and present the details in \cref{table:FixNonOrthCorr_Nnon}. 

As $N_{non}$ increases to $2$, both final errors are reduced by two orders of magnitude However, these results are still significantly outperformed by ResNonOrthCorr, whose CPU times are also significantly shorter. 

\begin{figure}[!htb]
     \begin{subfigure}[t]{.45\textwidth}
      \centering
      \includegraphics[width=\linewidth]{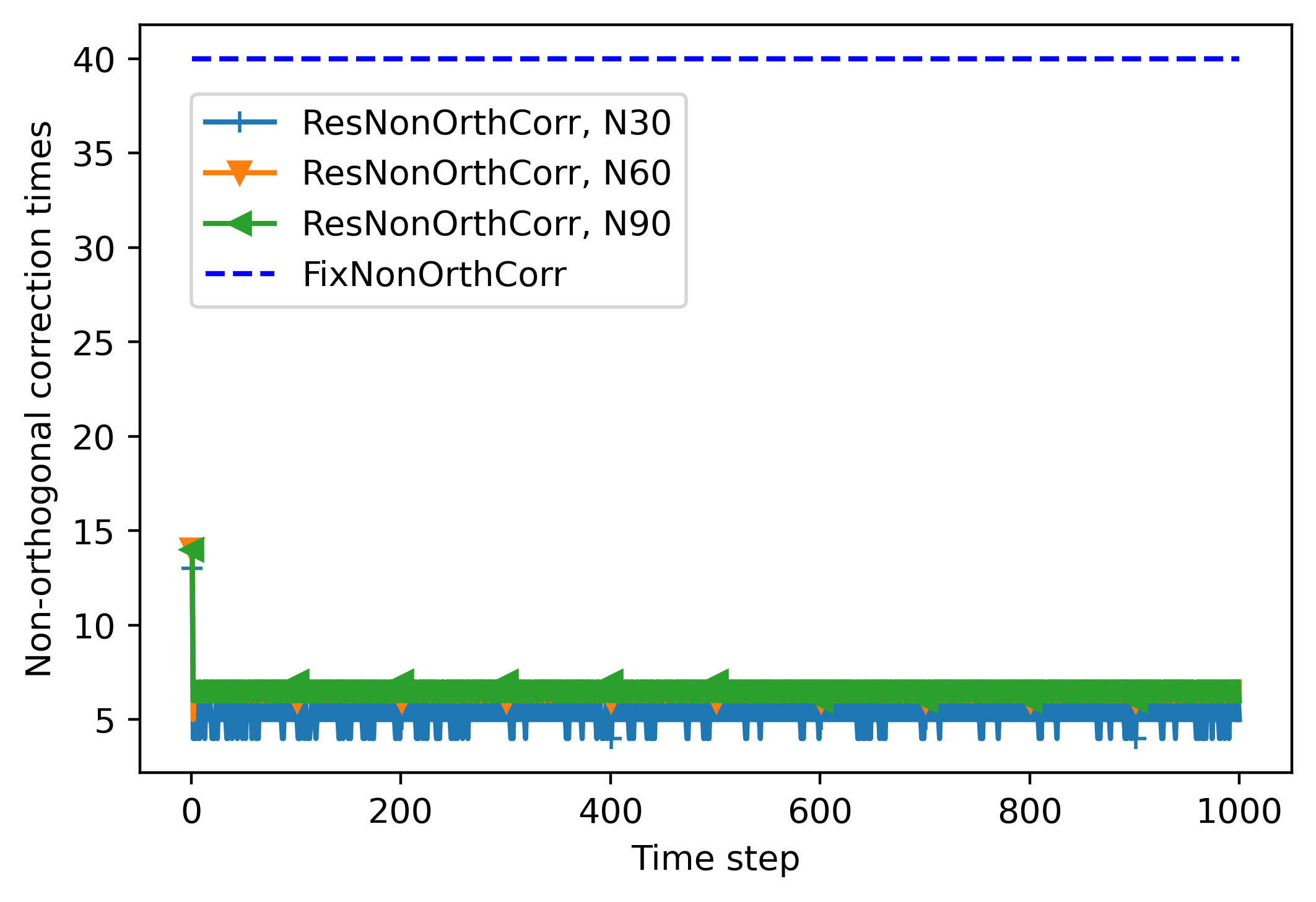}
      \caption{perturbMesh: temporal evolution of the number of non-orthogonality corrections.}
      \label{fig:staticDroplet3D_NnonTimes-evolution}
     \end{subfigure}
     \begin{subfigure}[t]{.45\textwidth}
      \centering
      \includegraphics[width=\linewidth]{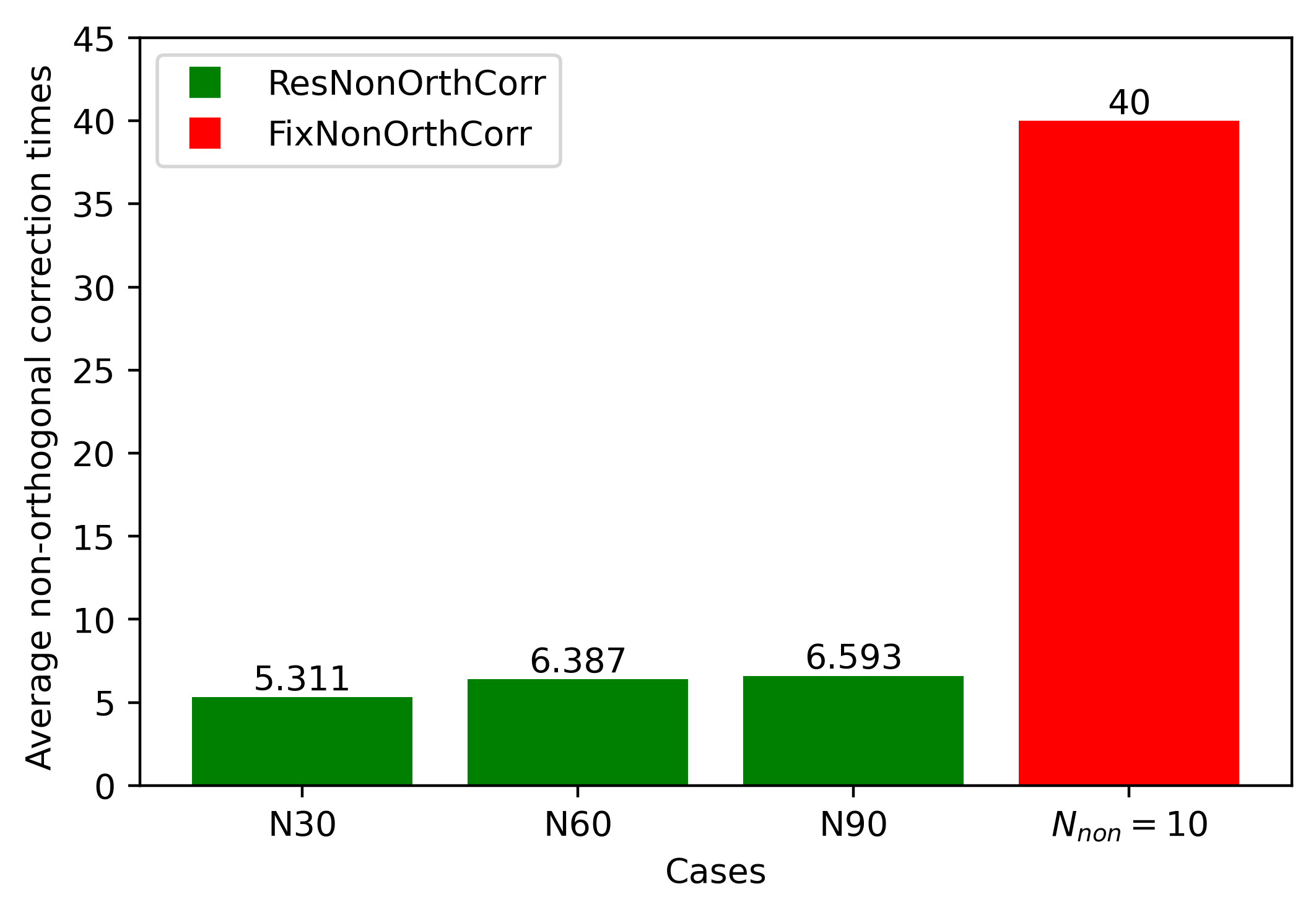}
      \caption{perturbMesh: average number of non-orthogonality corrections.}
      \label{fig:staticDroplet3D_NnonTimes-average}
     \end{subfigure}  
     \begin{subfigure}[t]{.5\textwidth}
      \centering
      \includegraphics[width=\linewidth]{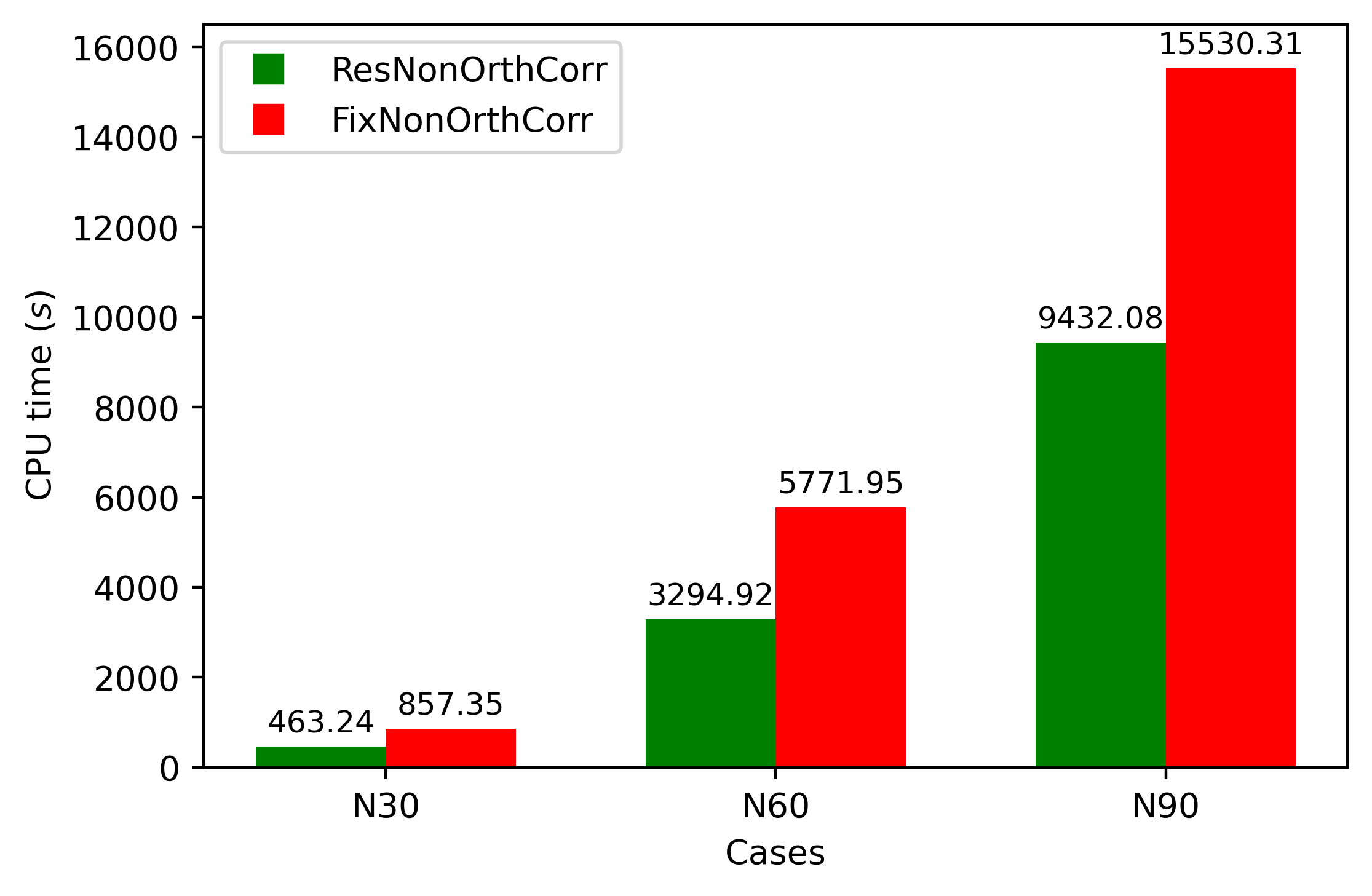}
      \caption{perturbMesh: CPU time; average speedup of ResNonOrthCorr $174.97\%$.}
      \label{fig:staticDroplet3D_CPUTime}
     \end{subfigure}
    \caption{ The temporal evolution and average of non-orthogonal correction times, and the CPU time with ResNonOrthCorr and FixNonOrthCorr ($N_{non}= 10$) for a stationary droplet in equilibrium on perturbMesh with different resolutions.}
    \label{fig:staticDroplet3D_NnonTimes}
\end{figure}

\begin{figure}[!htb]
     \begin{subfigure}{.45\textwidth}
      \centering
      \includegraphics[width=\linewidth]{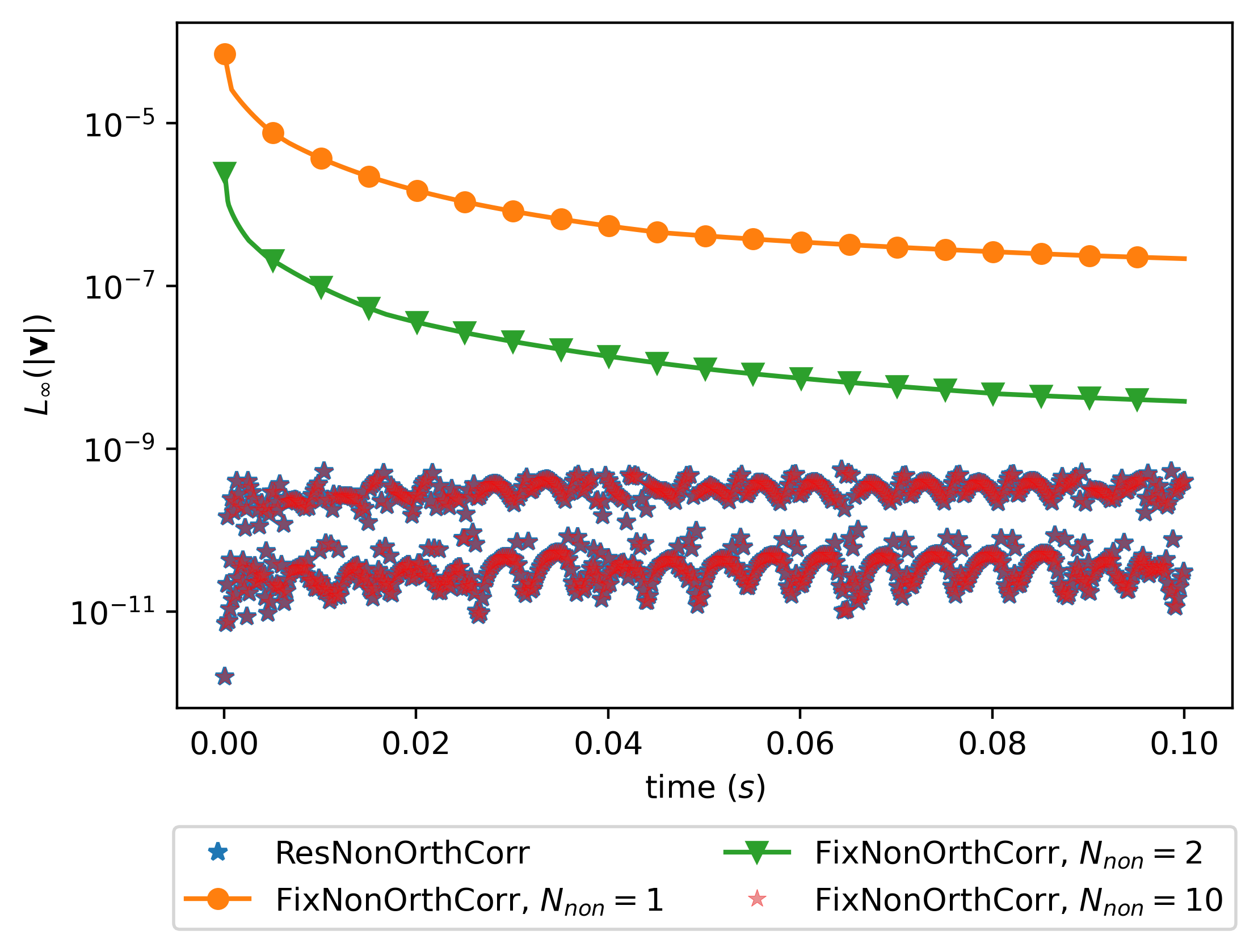}
      \caption{perturbMesh: $\frac{\Delta x}{L} = 1/30$}
     \end{subfigure}
     \begin{subfigure}{.45\textwidth}
      \centering
      \includegraphics[width=\linewidth]{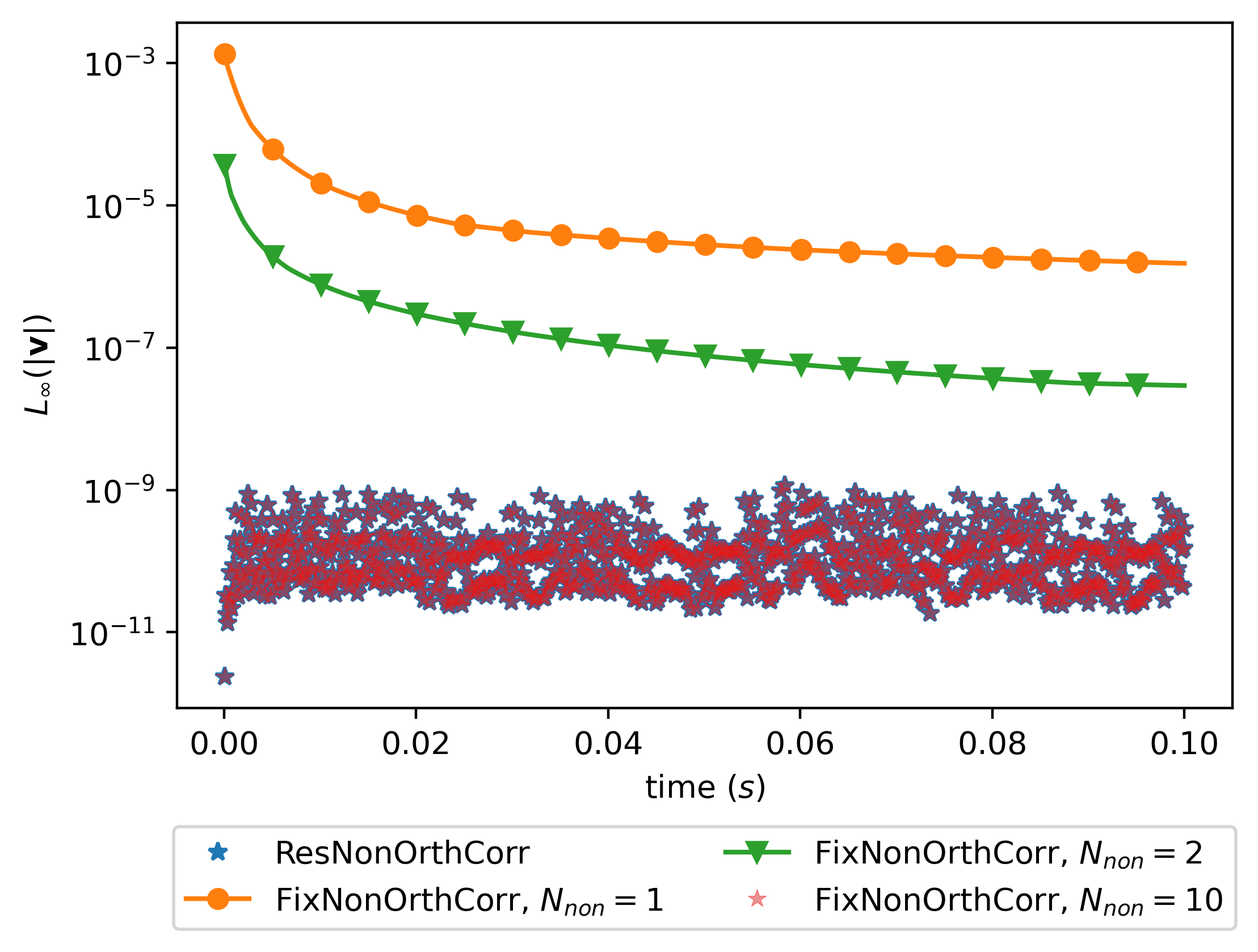}
      \caption{perturbMesh: $\frac{\Delta x}{L} = 1/60$}
     \end{subfigure}
     \begin{subfigure}{.45\textwidth}
      \centering
      \includegraphics[width=\linewidth]{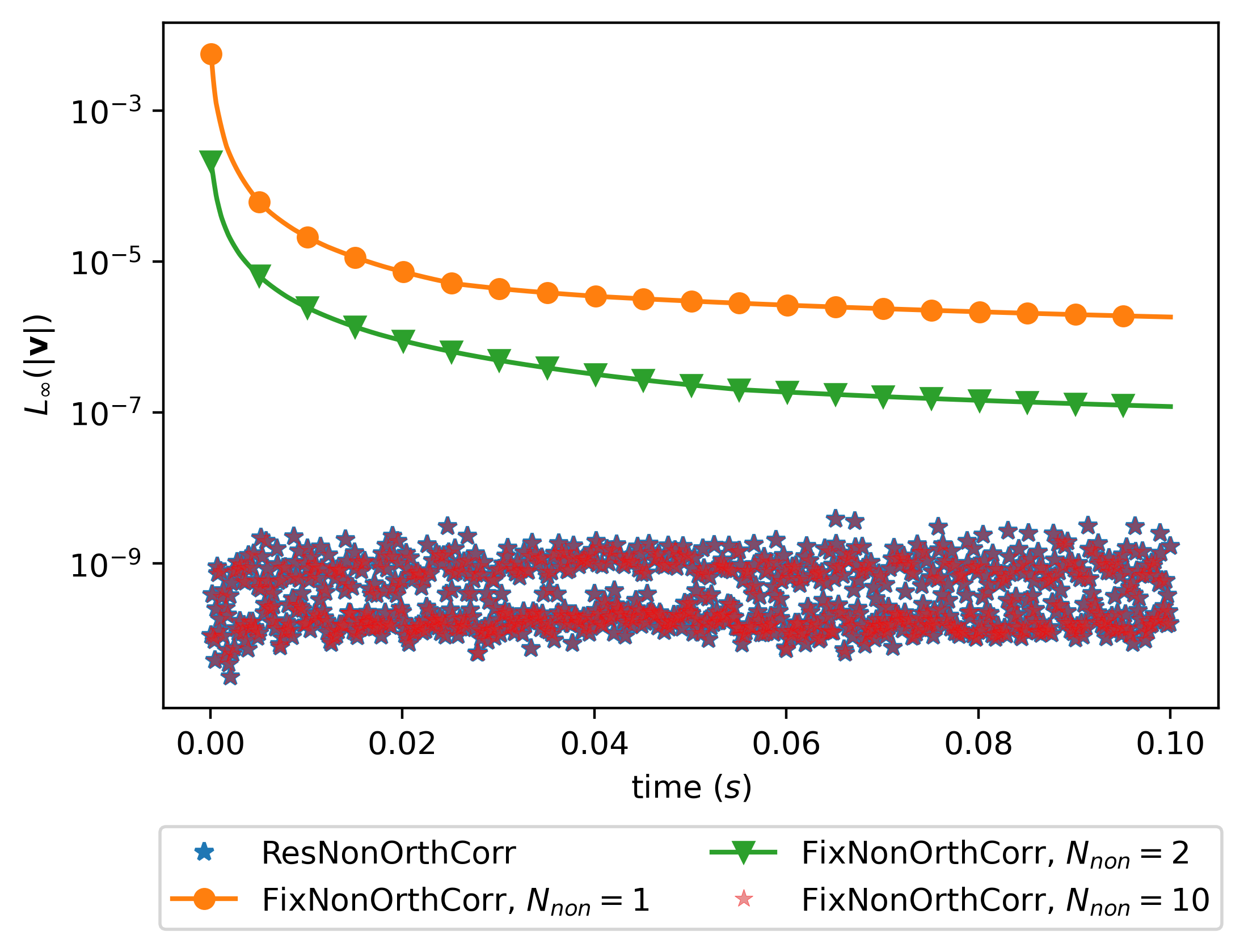}
      \caption{perturbMesh: $\frac{\Delta x}{L} = 1/90$}
     \end{subfigure}    
    \caption{ The temporal evolution of velocity error norm $L_{\infty}(|\v|)$ with ResNonOrthCorr and FixNonOrthCorr ($N_{non}= [1,2,10]$) for a stationary droplet in equilibrium on perturbMesh with different resolutions.}
    \label{fig:staticDroplet3D_perturbMesh}
\end{figure}

\begin{figure}[!htb]
     \begin{subfigure}{.45\textwidth}
      \centering
      \includegraphics[width=\linewidth]{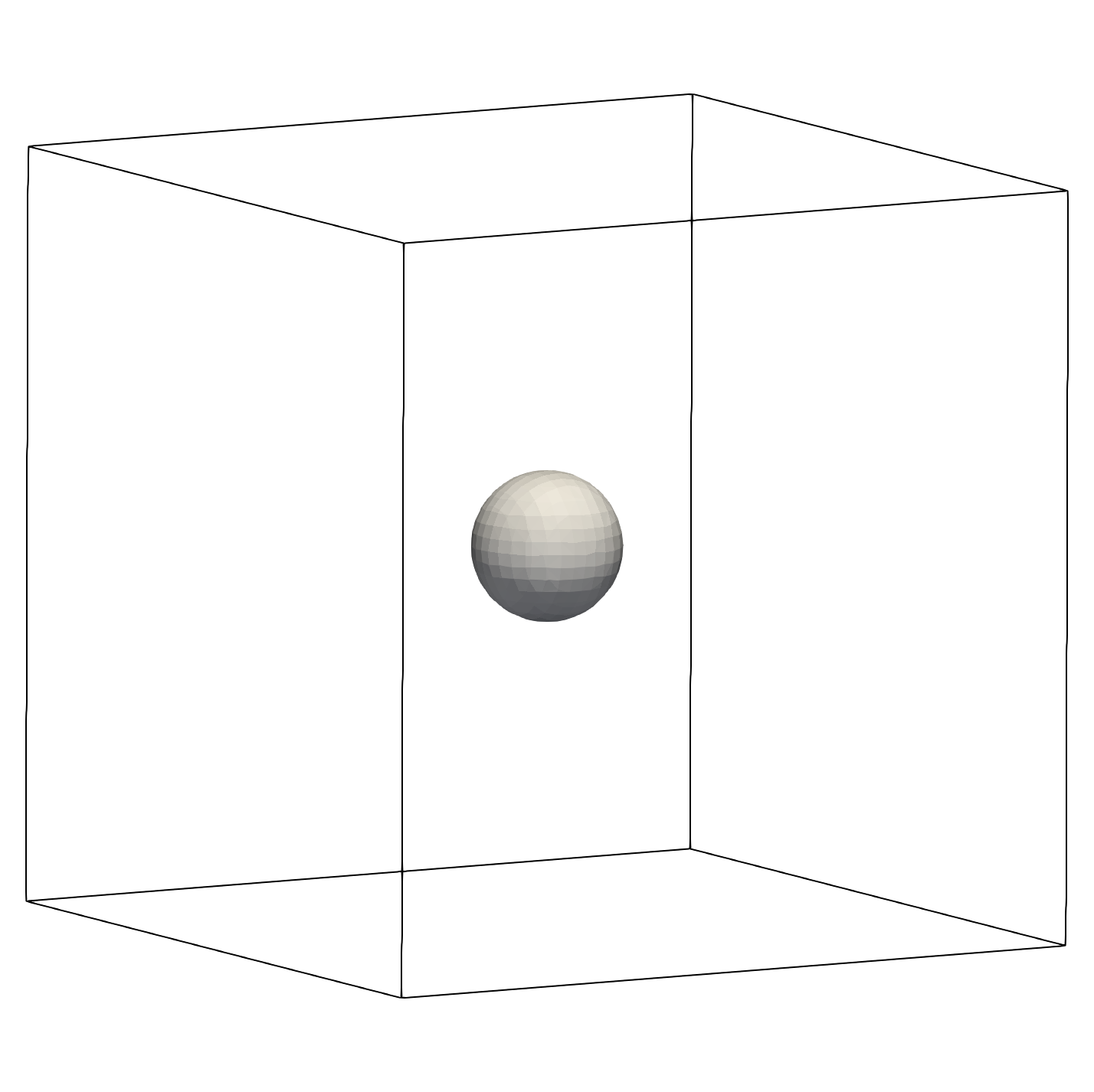}
      \caption{ResNonOrthCorr}
     \end{subfigure}
     \begin{subfigure}{.45\textwidth}
      \centering
      \includegraphics[width=\linewidth]{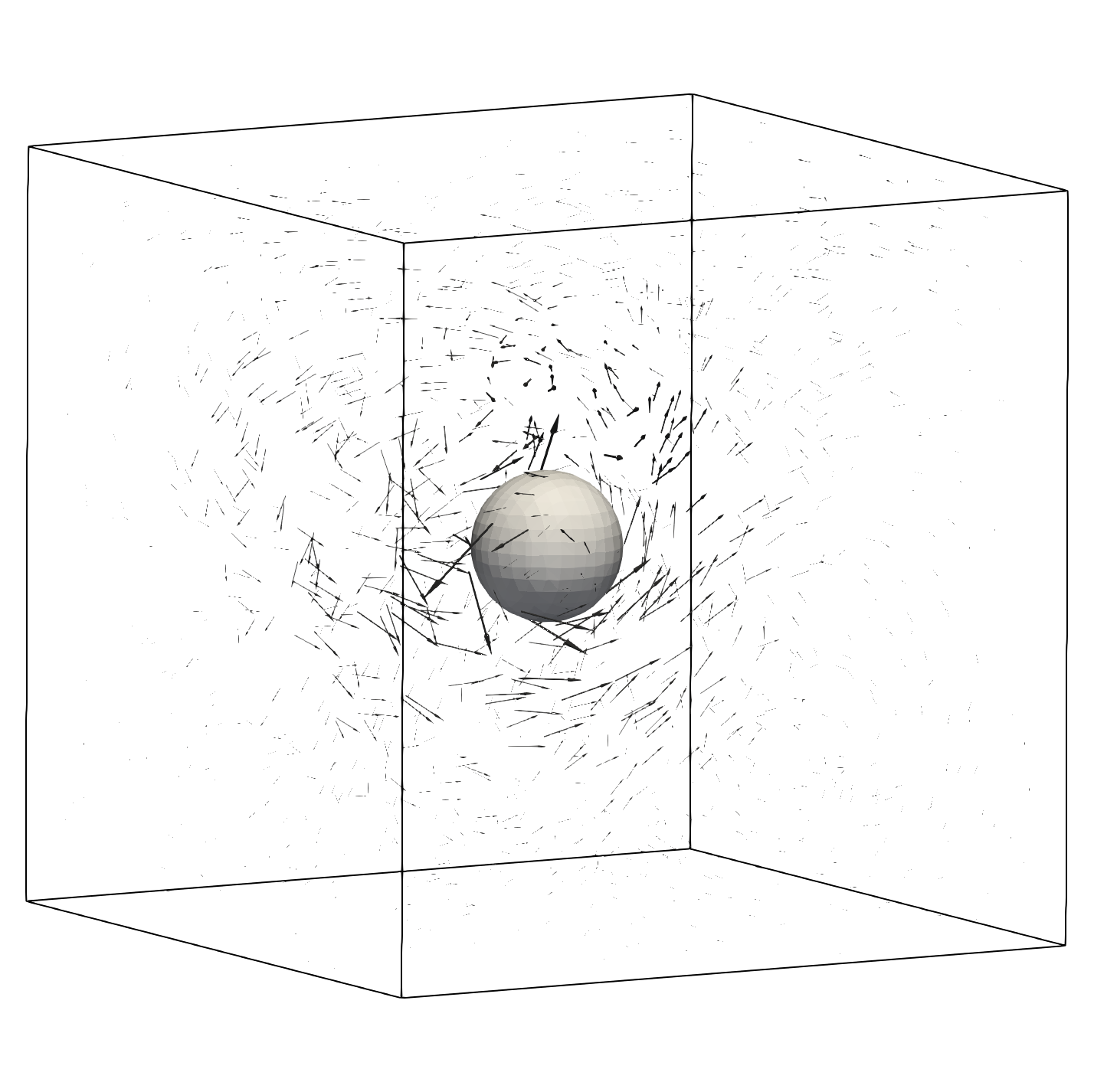}
      \caption{FixNonOrthCorr($N_{non}=1$)}
     \end{subfigure}   
    \caption{ The velocity field of the stationary droplet in equilibrium on perturbMesh with the resolution $\frac{\Delta x}{L} = 1/60$ at $t=t_{end}$: the glyph arrows are scaled by $1\text{e-}7\unit{m/s}$.}
    \label{fig:staticDroplet3D_perturbMesh_N60_maxU_glyph}
\end{figure}

When the non-orthogonal contribution is corrected $10$ times in each inner loop, i.e., $N_{non}=10$, for FixNonOrthCorr, the final errors exhibit minor differences compared to ResNonOrthCorr and approach the prescribed tolerance, suggesting that setting $N_{non}=10$ for this static droplet test is sufficient to mitigate errors attributable to non-orthogonality to a satisfactory level. However, the CPU time required to achieve a similar level of accuracy with FixNonOrthCorr and $N_{non}=10$ is approximately $70\%$ higher than that with ResNonOrthCorr for each resolution.  \Cref{fig:staticDroplet3D_NnonTimes} illustrates the temporal evolution and average non-orthogonal correction times. After sufficient corrections at the first time step, the non-orthogonality is corrected only around $5$ times for ResNonOrthCorr with all resolutions, as shown in \cref{fig:staticDroplet3D_NnonTimes-evolution}. Owing to the prescribed $N_{outer}=4$, $N_{inner}=1$ and $N_{non}=10$, the correction times for FixNonOrthCorr are fixed accordingly, i.e., $N_{outer}\times N_{inner} \times N_{non} = 40$, which explains the horizontal blue dashed line of FixNonOrthCorr in \cref{fig:staticDroplet3D_NnonTimes-evolution}. \Cref{fig:staticDroplet3D_NnonTimes-average} presents the average non-orthogonal correction times per time step regarding \cref{fig:staticDroplet3D_NnonTimes-evolution}. The correction times rise slightly as the resolution increases for ResNonOrthCorr, whereas the correction times are unchanged for FixNonOrthCorr regardless of the resolution.   
The \cref{fig:staticDroplet3D_perturbMesh} visually depicts the temporal evolution of $L_{\infty}(|\v|)$ with different control methods and indicates a trend: the errors from FixNonOrthCorr decreases as $N_{non}$ rises in the context of the non-orthogonal perturbMesh. When the specified $N_{non}$ is sufficiently large, such as $N_{non}=10$ in this case, FixNonOrthCorr works same as ResNonOrthCorr, as evidenced by the complete overlap of the blue stars from ResNonOrthCorr with the red transparent stars from FixNonOrthCorr with $N_{non}=10$ in \cref{fig:staticDroplet3D_perturbMesh}. Finally, \cref{fig:staticDroplet3D_perturbMesh_N60_maxU_glyph} visually represents the final velocity field on perturbMesh with ResNonOrthCorr and FixNonOrthCorr, where the orientation and length of each glyph arrow signify the direction and magnitude scaled by a factor of $1\text{e-}7$ of velocity at a cell center.

In addition to the prescribed Green Gauss method, we also explored a least-square method named \textit{pointCellsLeastSquares}. This method utilizes point-neighbor cells as the stencil to calculate the gradient, allowing us to investigate the impact of different cell center gradient reconstruction methods on non-orthogonality correction. As depicted in \cref{table:least-square_Nnon}, the errors obtained using the higher-order accurate least-square method are reduced at the final time for both ResNonOrthCorr and FixNonOrthCorr ($N_{non} = 10$). However, the more complex calculation involved in the least-square gradient results in higher computational costs, as evidenced by the increased CPU time shown in \cref{table:least-square_Nnon}. The temporal evolution of velocity errors for different gradient reconstructions is illustrated in \cref{fig:stationaryDroplet_perturbMesh_LeastSquare}.

\Cref{table:MULES_Nnon} presents the final results of errors and CPU time on perturbMesh using the Multidimensional universal limiter for explicit solution (MULES), another Volume of Fluid (VoF)-based interface capturing method in \OF \citep{greenshields2023}. In contrast to the geometric isoAdvector \citep{Scheufler2021} utilized earlier, MULES is an algebraic VoF solver that employs the Flux Corrected Transport (FCT) technique \citep{Boris1973flux,Zalesak1979fully}. We combined our ResNonOrthCorr with MULES. As shown in \cref{table:MULES_Nnon}, the errors obtained from MULES + ResNonOrthCorr approach the preset tolerance. On the other hand, the CPU time for all resolutions is very similar to the preceding results obtained from isoAdvector + ResNonOrthCorr. Our method demonstrates excellent compatibility with different phase advection methods. \Cref{fig:stationaryDroplet_perturbMesh_MULES} illustrates the velocity errors over time for MULES + ResNonOrthCorr. The combination of ResNonOrthCorr with MULES yields even better results for this stationary droplet case.
\begin{table}[!htb]
\begin{adjustbox}{width=.88\textwidth}
\small
\begingroup
\begin{tabular}{llllllll}
\toprule
Mesh type   & Resolution & \multicolumn{3}{l}{FixNonOrthCorr($N_{non}=2$)}     & \multicolumn{3}{l}{FixNonOrthCorr($N_{non}=10$)}                         \\ 
            & $\frac{\Delta x}{L}$      & \begin{tabular}[c]{@{}l@{}}$L_{\infty}(|\v|)$\\ ($\unit{m/s}$)\end{tabular} & \begin{tabular}[c]{@{}l@{}}$L(|\Delta p|)$\\ ($\unit{Pa}$)\end{tabular} & \begin{tabular}[c]{@{}l@{}}CPU time\\ ($s$)\end{tabular} & \begin{tabular}[c]{@{}l@{}}$L_{\infty}(|\v|)$\\ ($\unit{m/s}$)\end{tabular} & \begin{tabular}[c]{@{}l@{}}$L(|\Delta p|)$\\ ($\unit{Pa}$)\end{tabular} & \begin{tabular}[c]{@{}l@{}}CPU time\\ ($s$)\end{tabular}    \\ \midrule
perturbMesh & 1/30       & 3.81861e-09 & 8.7535e-12 & \multicolumn{1}{l|}{527.45}   & 4.0199e-10 & 3.3798e-13 & 857.35   \\
            & 1/60       & 2.9294e-08  & 8.5609e-11 & \multicolumn{1}{l|}{4947.63}  & 2.8729e-10 & 1.3269e-12 & 5771.95  \\
            & 1/90       & 1.1993e-07  & 2.6584e-10 & \multicolumn{1}{l|}{16894.48} & 1.7006e-9 & 2.2078e-12 & 15530.31 \\ \bottomrule
\end{tabular}
\endgroup
\end{adjustbox}
\caption{The performances of FixNonOrthCorr with non-orthogonality loop numbers larger than one for a stationary droplet in equilibrium case on perturbMesh.}
\label{table:FixNonOrthCorr_Nnon}
\end{table}

\begin{table}[!htb]
\begin{adjustbox}{width=.88\textwidth}
\small
\begingroup
\begin{tabular}{llllllll}
\toprule
Mesh type   & Resolution & \multicolumn{3}{l}{ResNonOrthCorr}     & \multicolumn{3}{l}{FixNonOrthCorr($N_{non}=10$)}                         \\ 
            & $\frac{\Delta x}{L}$      & \begin{tabular}[c]{@{}l@{}}$L_{\infty}(|\v|)$\\ ($\unit{m/s}$)\end{tabular} & \begin{tabular}[c]{@{}l@{}}$L(|\Delta p|)$\\ ($\unit{Pa}$)\end{tabular} & \begin{tabular}[c]{@{}l@{}}CPU time\\ ($s$)\end{tabular} & \begin{tabular}[c]{@{}l@{}}$L_{\infty}(|\v|)$\\ ($\unit{m/s}$)\end{tabular} & \begin{tabular}[c]{@{}l@{}}$L(|\Delta p|)$\\ ($\unit{Pa}$)\end{tabular} & \begin{tabular}[c]{@{}l@{}}CPU time\\ ($s$)\end{tabular}    \\ \midrule
perturbMesh & 1/30       & 6.3041e-11 & 1.4322e-12 & \multicolumn{1}{l|}{663.04}   & 6.3041e-11 & 1.4322e-12 & 1066.53   \\
            & 1/60       & 6.5157e-11  & 7.1151e-13 & \multicolumn{1}{l|}{5166.65}  & 6.5157e-10 & 6.8416e-13 & 7171.11  \\
            & 1/90       & 4.0325e-11  & 5.7556e-12 & \multicolumn{1}{l|}{13342.54} & 4.0325e-11 & 5.7281e-12 & 19408.88 \\ \bottomrule
\end{tabular}
\endgroup
\end{adjustbox}
\caption{The performances of ResNonOrthCorr and FixNonOrthCorr($N_{non}=10$) with least-square cell center gradient reconstruction method for a stationary droplet in equilibrium case on perturbMesh.}
\label{table:least-square_Nnon}
\end{table}
\begin{table}[!htb]
\begin{adjustbox}{width=.55\textwidth}
\small
\begingroup
\begin{tabular}{lllll}
\toprule
Mesh type   & Resolution & \multicolumn{3}{l}{ResNonOrthCorr: MULES}                            \\ 
            & $\frac{\Delta x}{L}$      & \begin{tabular}[c]{@{}l@{}}$L_{\infty}(|\v|)$\\ ($\unit{m/s}$)\end{tabular} & \begin{tabular}[c]{@{}l@{}}$L(|\Delta p|)$\\ ($\unit{Pa}$)\end{tabular} & \begin{tabular}[c]{@{}l@{}}CPU time\\ ($s$)\end{tabular} \\ \midrule
perturbMesh & 1/30       & 6.1017e-12 & 5.6519e-13 & 371.47      \\
            & 1/60       & 9.6775e-11  & 1.2905e-11 & 3398.93    \\
            & 1/90       & 1.8768e-11  & 1.2631e-11 & 10096.52 \\ \bottomrule
\end{tabular}
\endgroup
\end{adjustbox}
\caption{The performances of combining MULES phase advection algorithm with ResNonOrthCorr for a stationary droplet in equilibrium case on perturbMesh.}
\label{table:MULES_Nnon}
\end{table}


In contrast to perturbMesh, the distribution of non-orthogonality within the polyMesh is irregular. As illustrated in \cref{table:staticDroplet3D}, the maximum local $\theta_f$ values are notably small, while the maximum global $\theta_f$ values exceed $30\unit{\degree}$ for all resolutions. This observation indicates that cells near the interface exhibit nearly orthogonal characteristics. Consequently, despite only using a single correction for non-orthogonality (i.e., once in each inner loop, $N_{non}=1$), the velocity and pressure jump outcomes from FixNonOrthCorr on polyMesh align with the high accuracy achieved on blockMesh. The substantial non-orthogonality does not necessitate a higher $N_{non}$, rendering the determination of an appropriate $N_{non}$ more arbitrary and intricate for users. ResNonOrthCorr, in contrast, circumvents this issue and yields satisfactory results.  

To conclude the results for the stationary droplet, the ResNonOrthCorr algorithm successfully ensures force-balance for the surface tension force and the pressure gradient on non-orthogonal unstructured finite volume discretization. It adjusts the number of required non-orthogonality iterations to satisfy \cref{eq:balancePressureErrorNorm}, a very cost-effective stopping criterion. The total CPU time is significantly reduced compared to the heuristic approach, since the number of non-orthogonal corrections is kept to a minimum. Finally, and equally impactful, ResNonOrthCorr removes the number of non-orthogonality corrections as a heuristic user-defined parameter from the CFD simulation. We have found in this verification study that $N_{non}=10$ ensures sufficient accuracy (at a much higher computational cost); however, finding this number for a highly resolved geometrically complex 3D microfluidic simulation is nearly impossible, as it is not possible to know the maximal interface-local non-orthogonality of all the cells that will be visited by the fluid interface during a simulation.

\subsection{Stationary water column in equilibrium}
In order to investigate the balance between gravitational force and the associated pressure gradient, we consider a stationary water column in equilibrium. In this analysis, the effects of surface tension force and viscosity are neglected. 
The test employs the same discretization scheme for density and pressure gradients to satisfy the fundamental force-balance requirement. It is crucial to note that inappropriately estimating $\mathbf{x}$ in $(\mathbf{g}\cdot\mathbf{x})$ can also deteriorate the force-balance \citep{Montazeri2010,Montazeri2012,Montazeri2014,Kasper2022}, particularly for the multiphase flows characterized by high density-ratios. This is because any error from the estimation of $\mathbf{x}$ is amplified by $(\rho^- - \rho^+)|\mathbf{g}|$, which approaches $10^4$ for water/air on earth. However, addressing the accuracy improvement of $\mathbf{x}$ estimation falls outside the scope of this study. Interested readers are directed to \citep{Montazeri2010,Kasper2022} for further details. For an inviscid water column in equilibrium, the interface $\Sigma$ between water and air is flat, resulting in a constant value for $(\mathbf{g}\cdot\mathbf{x}_{\Sigma})$.

The cubic container of the water column is used as the computation domain $\Omega$, with dimensions of $[0,0,0]\times[1,1,1]\,\unit{m}$. The water occupies the region where $z_{\Sigma}\le0.5145\,\unit{m}$ within $\Omega$. All the boundaries are treated as walls, except for the top boundary, which is modeled by the open-air boundary condition.
The relevant properties of water and air, as well as gravity, are defined in \cref{table:waterAirProperties}. The error norms from \cref{eq:error_norms} are employed, along with the exact solution
\begin{equation*}
  \begin{aligned}
        \v_e &= \mathbf{0}\ \unit{m/s} \\
        \Delta p_e &= (\rho^- - \rho^+)|\g|z_{\Sigma} = 5040.95633874\ \unit{Pa}.
    \end{aligned}
\end{equation*}

\begin{table}[]
\begin{adjustbox}{width=\textwidth}
\small
\begingroup
\begin{tabular}{llllllllll}
\toprule
Mesh type                    & Resolution          &  \multicolumn{2}{l}{max. non-ortho.}                                      & \multicolumn{3}{l}{ResNonOrthCorr}                                                                                                                                                                                       & \multicolumn{3}{l}{FixNonOrthCorr($N_{non}=1$)}\\
                             & $\frac{\Delta x}{L}$ &\begin{tabular}[c]{@{}l@{}}global($\theta_f$) \\ ($\unit{\degree}$)\end{tabular} &\begin{tabular}[c]{@{}l@{}}local($\theta_f$) \\ ($\unit{\degree}$)\end{tabular}     & \begin{tabular}[c]{@{}l@{}}$L_{\infty}(|\v|)$\\ ($\unit{m/s}$)\end{tabular} & \begin{tabular}[c]{@{}l@{}}$L(|\Delta p|)$\\ ($\unit{Pa}$)\end{tabular} & \begin{tabular}[c]{@{}l@{}}CPU time\\ ($s$)\end{tabular} & \begin{tabular}[c]{@{}l@{}}$L_{\infty}(|\v|)$\\ ($\unit{m/s}$)\end{tabular} & \begin{tabular}[c]{@{}l@{}}$L(|\Delta p|)$\\ ($\unit{Pa}$)\end{tabular} & \begin{tabular}[c]{@{}l@{}}CPU time\\ ($s$)\end{tabular} \\ \midrule
blockMesh   & 1/30       & 0                & 0                & 6.1977e-14 & 7.1573e-13 & \multicolumn{1}{l|}{686.11}   & 1.4301e-13 & 7.1898e-13 & 675.24   \\
            & 1/60       & 0                & 0                & 7.8625e-12 & 4.0783e-13 & \multicolumn{1}{l|}{4052.7}   & 7.8600e-12 & 4.0801e-13 & 4018.51  \\
            & 1/90       & 0                & 0                & 6.6016e-12 & 1.7352e-13 & \multicolumn{1}{l|}{10641.22} & 6.6131e-12 & 1.7316e-13 & 10587.81 \\ \midrule
perturbMesh & 1/30       & 12.79            & 12.16            & 1.3731e-10 & 6.0642e-13 & \multicolumn{1}{l|}{725.66}   & 3.5756e-05 & 2.7689e-11 & 725.8    \\
            & 1/60       & 14.45            & 13.64            & 4.9291e-11 & 3.3008e-13 & \multicolumn{1}{l|}{4154.27}  & 1.0977e-04 & 9.9943e-11 & 4349.29  \\
            & 1/90       & 13.54            & 12.77            & 1.7483e-11 & 5.9791e-12 & \multicolumn{1}{l|}{10872.64} & 6.7688e-05 & 3.0113e-11 & 11505.59 \\ \midrule
polyMesh    & 1/30       &  60.41          &  58.88            & 1.4677e-09 & 1.8047e-08 & \multicolumn{1}{l|}{3232.11}  & 7.5769e-03 & 8.5541e-08 & 3516.54  \\
            & 1/60       & 63.30            & 58.98             & 1.1224e-08 & 2.4435e-09 & \multicolumn{1}{l|}{16538.47} & 3.3555e-02 & 2.7190e-07 & 21090.88 \\
            & 1/90       & 60.64            & 58.73            & 4.5175e-08 & 2.8148e-08 & \multicolumn{1}{l|}{49231.96} & 8.2165e-03 & 6.1529e-08 & 57566.64 \\ \bottomrule
\end{tabular}
\endgroup
\end{adjustbox}
\caption{The maximum non-orthogonalities in the global region and the local region near the interface, the velocity, pressure jump errors, and CPU time for a stationary water column in equilibrium at the end time $t_{end}=0.1s$}
\label{table:waterColumnResults}

\end{table}

Similar to the stationary droplet case, when non-orthogonality is uniformly zero across the entire computational domain for blockMesh, ResNonOrthCorr and FixNonOrthCorr ($N_{non}=1$) exhibit closely comparable performance in terms of velocity and pressure jump errors, as well as CPU time, presented in \cref{table:waterColumnResults}. 

The maximal local non-orthogonality exceeds $\ang{10}$ for perturbMesh, leading to noticeable disparities in errors between ResNonOrthCorr and FixNonOrthCorr. Specifically, with regard to the velocity field, errors with ResNonOrthCorr are $5$ to $7$ orders of magnitude smaller than those observed with FixNonOrthCorr. Although the CPU times for simulations with ResNonOrthCorr control are slightly lower than those using FixNonOrthCorr, the error discrepancies are significant. 

The polyMesh for this test shows the highest local non-orthogonality when compared with blockMesh and perturbMesh. Although the mesh generation is the same as for \cref{table:staticDroplet3D}, here the fluid interface touches the domain boundary, that has maximal non-orthogonality.  Despite the very large non-orthogonality of $\approx60^\circ$, ResNonOrthCorr effectively achieves force balance with velocity error magnitudes of $\approx1\text{e-}8$. In contrast, parasitic velocities arising from the use of FixNonOrthCorr reach the magnitude of $1\text{e-}2$. The CPU times of ResNonOrthCorr are additionally comparatively lower than those of FixNonOrthCorr.   

\begin{figure}[htb]
     \begin{subfigure}{.4\textwidth}
      \centering
      \includegraphics[width=\linewidth]{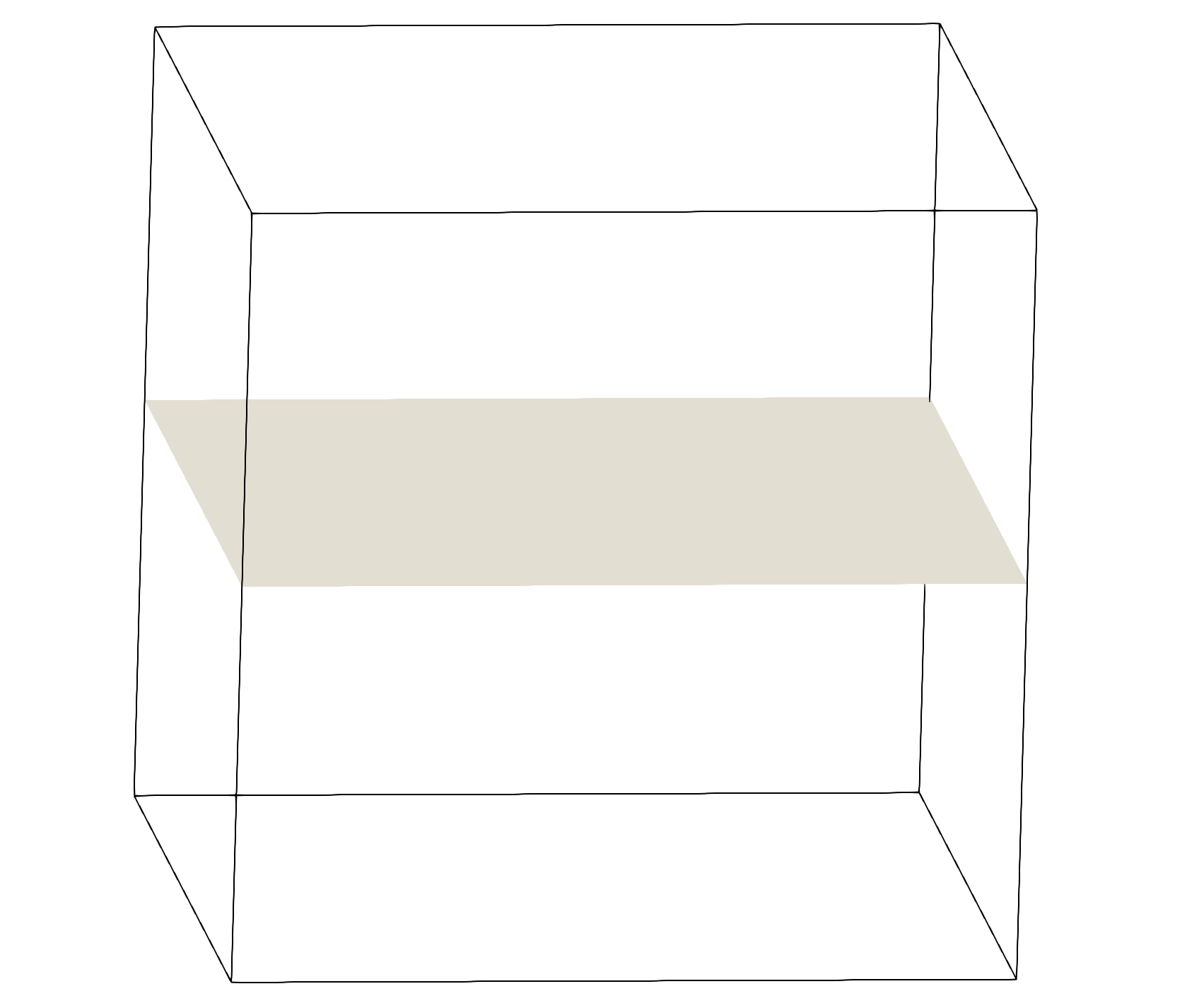}
      \caption{ResNonOrthCorr}
     \end{subfigure}
     \begin{subfigure}{.4\textwidth}
      \centering
      \includegraphics[width=\linewidth]{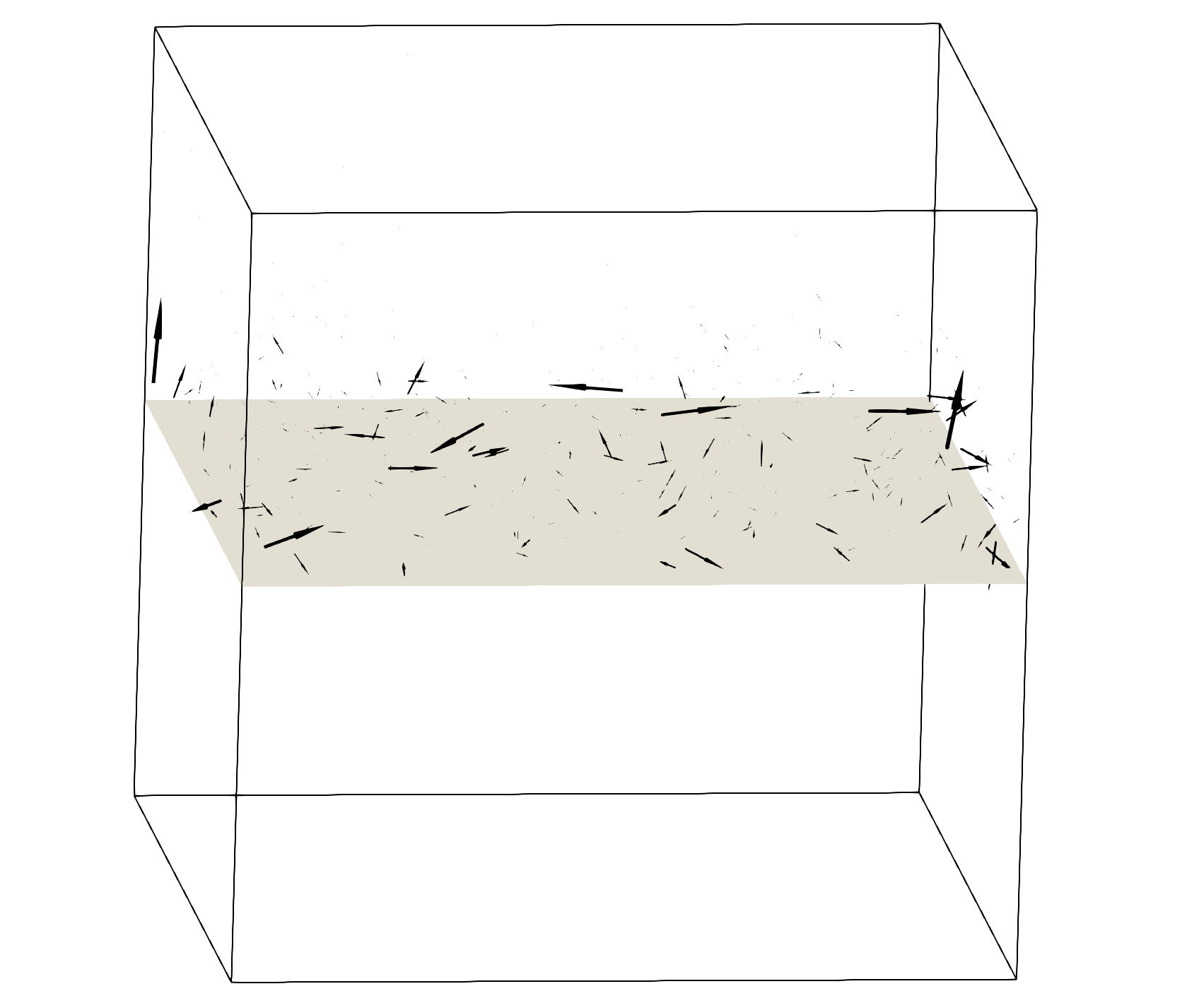}
      \caption{FixNonOrthCorr($N_{non}=1$)}
     \end{subfigure}   
    \caption{ The velocity field of the stationary water column in equilibrium on perturbMesh with the resolution $\frac{\Delta x}{L} = 1/60$ at $t=t_{end}$: the glyph arrows are scaled by $1\text{e-}5\unit{m/s}$.}
    \label{fig:waterColumn_perturbMesh_N60_maxU_glyph}
\end{figure}

\begin{table}[]
\begin{adjustbox}{width=.88\textwidth}
\small
\begingroup
\begin{tabular}{llllllll}
\toprule
Mesh type   & Resolution & \multicolumn{3}{l}{FixNonOrthCorr($N_{non}=2$)}     & \multicolumn{3}{l}{FixNonOrthCorr($N_{non}=10$)}                         \\ 
            & $\frac{\Delta x}{L}$      & \begin{tabular}[c]{@{}l@{}}$L_{\infty}(|\v|)$\\ ($\unit{m/s}$)\end{tabular} & \begin{tabular}[c]{@{}l@{}}$L(|\Delta p|)$\\ ($\unit{Pa}$)\end{tabular} & \begin{tabular}[c]{@{}l@{}}CPU time\\ ($s$)\end{tabular} & \begin{tabular}[c]{@{}l@{}}$L_{\infty}(|\v|)$\\ ($\unit{m/s}$)\end{tabular} & \begin{tabular}[c]{@{}l@{}}$L(|\Delta p|)$\\ ($\unit{Pa}$)\end{tabular} & \begin{tabular}[c]{@{}l@{}}CPU time\\ ($s$)\end{tabular}    \\ \midrule
perturbMesh & 1/30       & 8.0268e-10 & 1.8642e-11 & \multicolumn{1}{l|}{748.68}   & 1.3731e-10 & 1.9503e-10 & 965.61   \\
            & 1/60       & 1.5662e-09 & 1.4149e-10 & \multicolumn{1}{l|}{4321.92}  & 4.9291e-11 & 1.1832e-10 & 5989.96  \\
            & 1/90       & 1.1303e-09 & 1.4348e-10 & \multicolumn{1}{l|}{11422.03} & 1.7483e-11 & 1.6857e-09 & 16392.02 \\ \midrule
polyMesh    & 1/30       & 1.8107e-04 & 2.2694e-09 & \multicolumn{1}{l|}{3466.42}  & 1.7267e-10 & 3.0471e-08 & 4447.46  \\
            & 1/60       & 9.5487e-04 & 4.0979e-09 & \multicolumn{1}{l|}{18902.18} & 1.0068e-08 & 1.6252e-09 & 24490.22 \\
            & 1/90       & 7.5324e-05 & 5.4888e-10 & \multicolumn{1}{l|}{54671.07} & 1.1127e-08 & 8.1320e-09 & 76854.99 \\ \bottomrule
\end{tabular}
\endgroup
\end{adjustbox}
\caption{The performances of FixNonOrthCorr with non-orthogonality loop numbers larger than one for a stationary water column
in equilibrium case on perturbMesh and polyMesh.}
\label{table:waterColumnFixNonOrthCorr}
\end{table}
\Cref{table:waterColumnFixNonOrthCorr} contains the errors and CPU times from FixNonOrthCorr with $N_{non} > 1$ on both perturbMesh and polyMesh. Clearly, the errors decrease with increasing $N_{non}$, but significantly more CPU time is required as $N_{non}$ increases for each resolution on both meshes. \Cref{fig:waterColumn_perturbMesh} and \Cref{fig:waterColumn_polyMesh} illustrate the temporal evolution of the velocity error norm $L_{\infty}(|\v|)$ with ResNonOrthCorr and FixNonOrthCorr ($N_{non}= [1,2,10]$) for a stationary water column in equilibrium on perturbMesh and polyMesh respectively. After correcting the explicit non-orthogonal part of the face gradient a sufficient number of times, e.g., $N_{non} = 10$ for perturbMesh, the final errors in velocity converge to the results obtained with ResNonOrthCorr as depicted in \cref{fig:waterColumn_perturbMesh}. For the polyMesh owning the largest non-orthogonality, for $N_{non} = 10$, FixNonOrthCorr still cannot recover force balance as accurately as ResNonOrthCorr do, as illustrated in \cref{fig:waterColumn_polyMesh}, where the red star symbols from FixNonOrthCorr with $N_{non} = 10$ are still slightly higher than the blue stars from ResNonOrthCorr. The significantly increased CPU time associated with the higher $N_{non}$, as shown in \cref{table:waterColumnResults,table:waterColumnFixNonOrthCorr}, implies the poor balance between the result accuracy and the computational costs for FixNonOrthCorr method. On the contrary, ResNonOrthCorr can achieve high accuracy with relatively much lower computational costs. 
\Cref{fig:waterColumn_perturbMesh_N60_maxU_glyph} shows the final velocity field using ResNonOrthCorr and FixNonOrthCorr on perturbMesh with the middle mesh resolution. The parasitic currents are distributed randomly above the flat interface in the results using FixNonOrthCorr, whereas they cannot be observed at the same scale when adopting ResNonOrthCorr.


\section{Conclusions}
\label{sec:concl}

We propose a highly accurate, deterministic, and computationally efficient residual-based control for the non-orthogonality correction in the unstructured Finite Volume method, which balances the pressure gradient, gravity force, and the surface tension force at fluid interfaces using the solver's residual norm up to the tolerance of a linear solver. The proposed control ensures an accurate force balance without incurring a noticeable computational overhead even on highly challenging randomly perturbed hexahedral meshes that do not allow artificial error cancellation. Using the linear solver's norm and tolerance in a deterministic stopping criterion sets the minimal number of non-orthogonality corrections for maximal accuracy, contrary to approaches that require modification of the solution algorithm or select the number of non-orthogonality corrections using trial and error. In addition to a very high degree of accuracy, results confirm an order-of-magnitude increase in computational efficiency compared to the heuristic approach.

\section{Acknowledgments}

The last author acknowledges funding by the German Research Foundation (DFG)
– Project-ID 265191195 – SFB 1194.

Calculations for this research were conducted on the Lichtenberg high performance computer of the TU Darmstadt. 

\newpage
\appendix
\section{Supplementary results for Stationary droplet in equilibrium}
\label{Appendix:figures_stationaryDroplet}
\begin{figure}[htb]
     \begin{subfigure}{.45\textwidth}
      \centering
      \includegraphics[width=\linewidth]{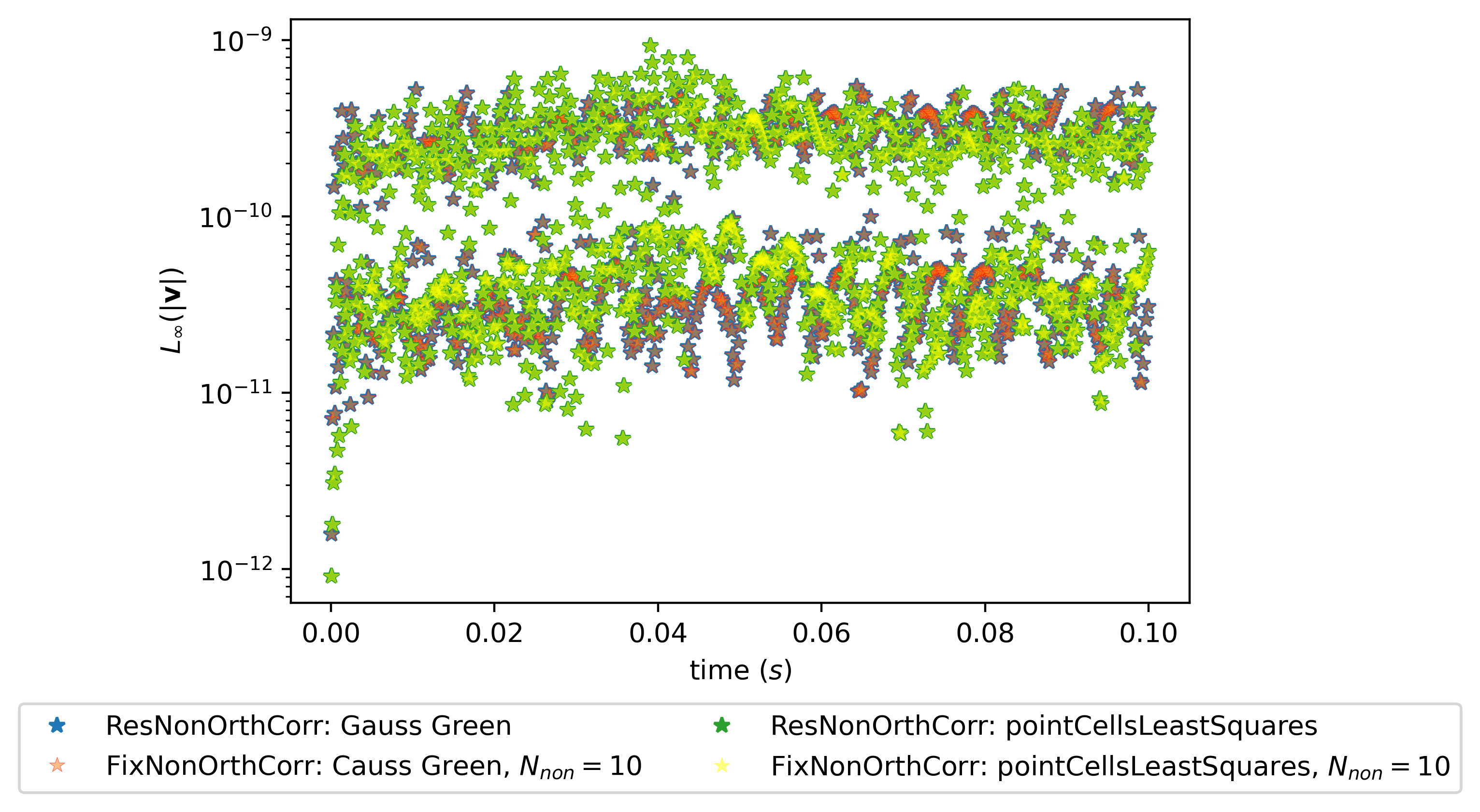}
      \caption{perturbMesh: $\frac{\Delta x}{L} = 1/30$}
     \end{subfigure}
     \begin{subfigure}{.45\textwidth}
      \centering
      \includegraphics[width=\linewidth]{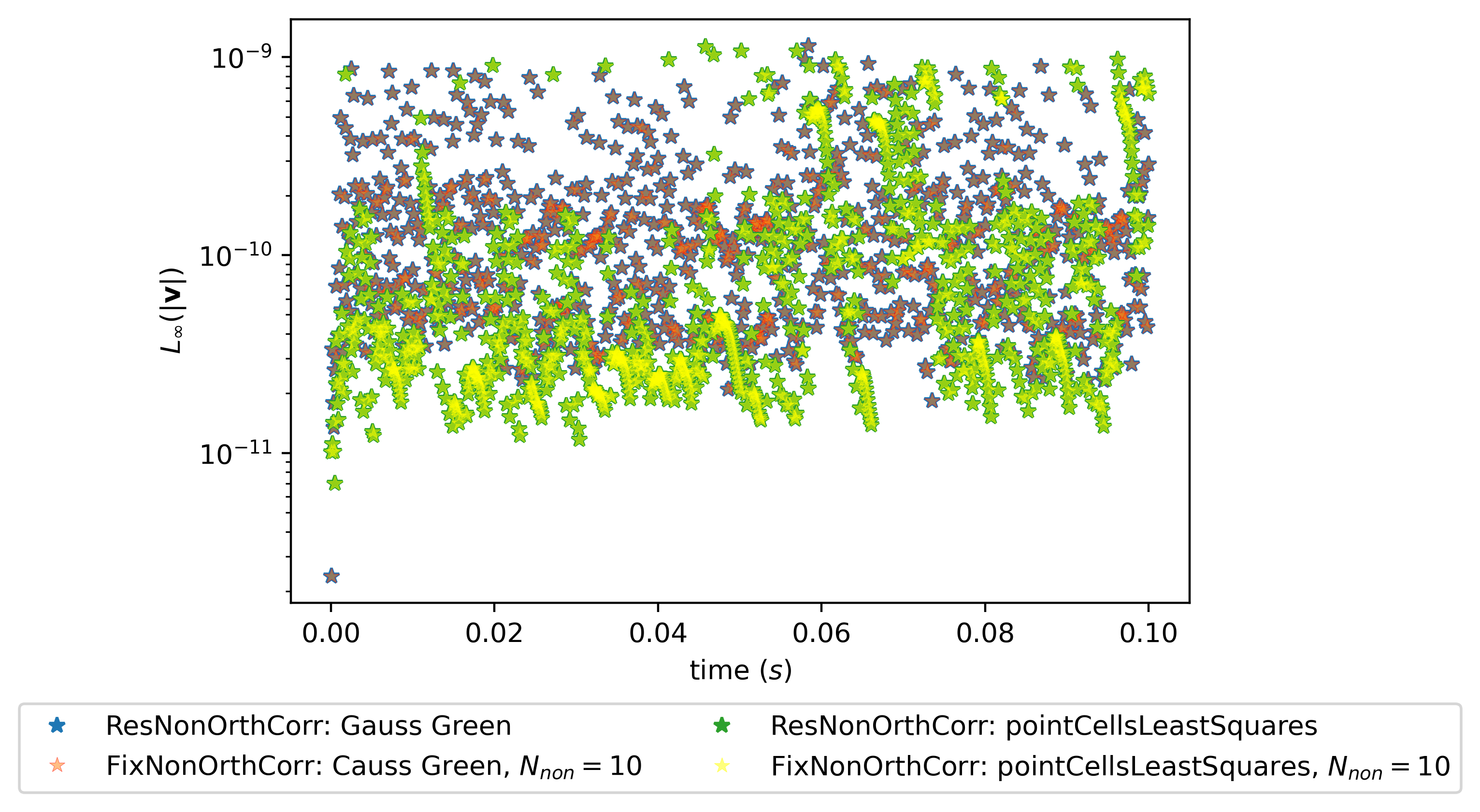}
      \caption{perturbMesh: $\frac{\Delta x}{L} = 1/60$}
     \end{subfigure}
     \begin{subfigure}{.45\textwidth}
      \centering
      \includegraphics[width=\linewidth]{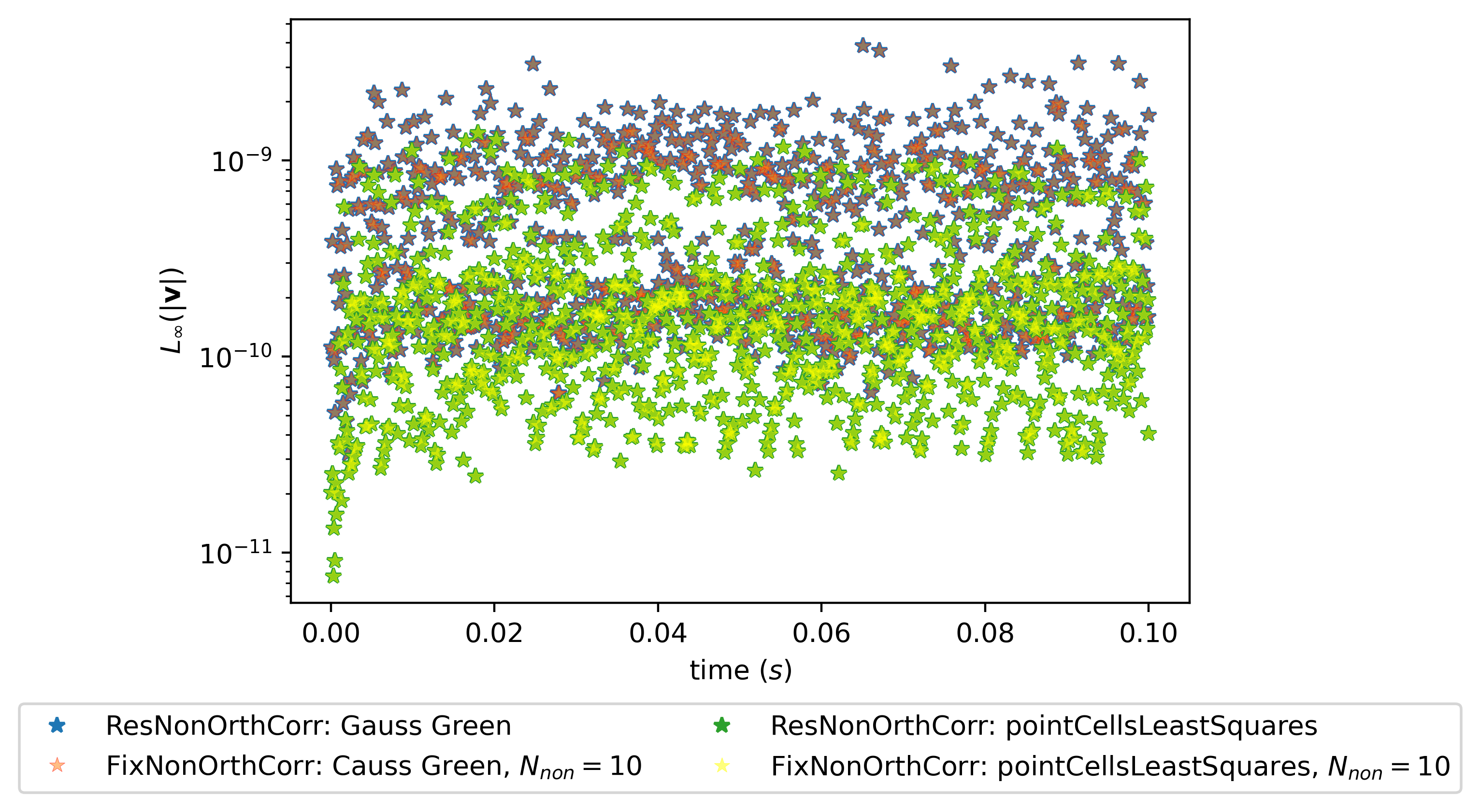}
      \caption{perturbMesh: $\frac{\Delta x}{L} = 1/90$}
     \end{subfigure}    
    \caption{The temporal evolution of velocity error norm $L_{\infty}(|\v|)$ using ResNonOrthCorr and FixNonOrthCorr ($N_{non}= 10$) with least-square gradient reconstruction for a stationary droplet in equilibrium on perturbMesh with different resolutions. These results show that there is no  influence on the gradient scheme on the proposed method.}
    \label{fig:stationaryDroplet_perturbMesh_LeastSquare}
\end{figure}

\begin{figure}[htb]
     \begin{subfigure}{.45\textwidth}
      \centering
      \includegraphics[width=\linewidth]{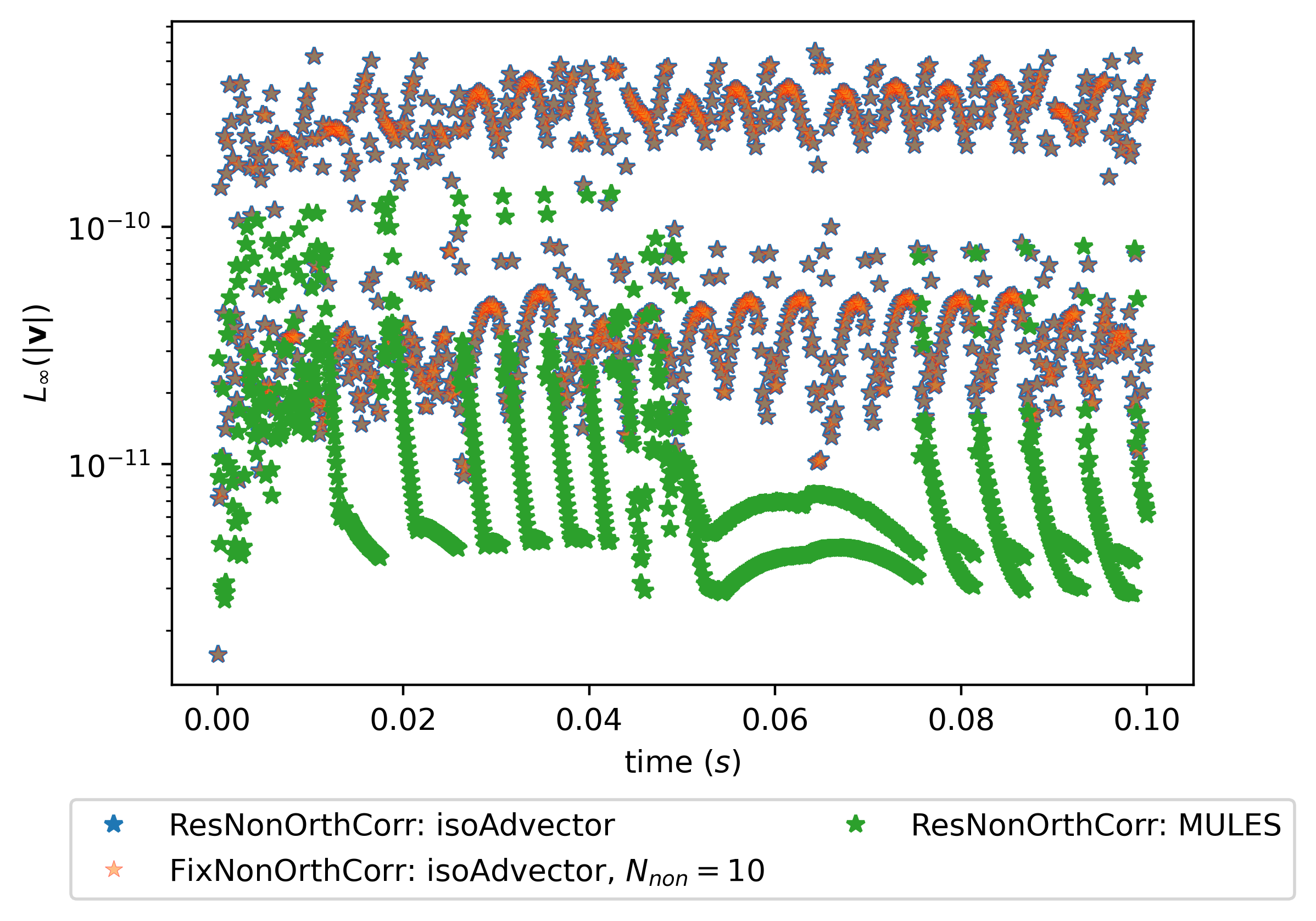}
      \caption{perturbMesh: $\frac{\Delta x}{L} = 1/30$}
     \end{subfigure}
     \begin{subfigure}{.45\textwidth}
      \centering
      \includegraphics[width=\linewidth]{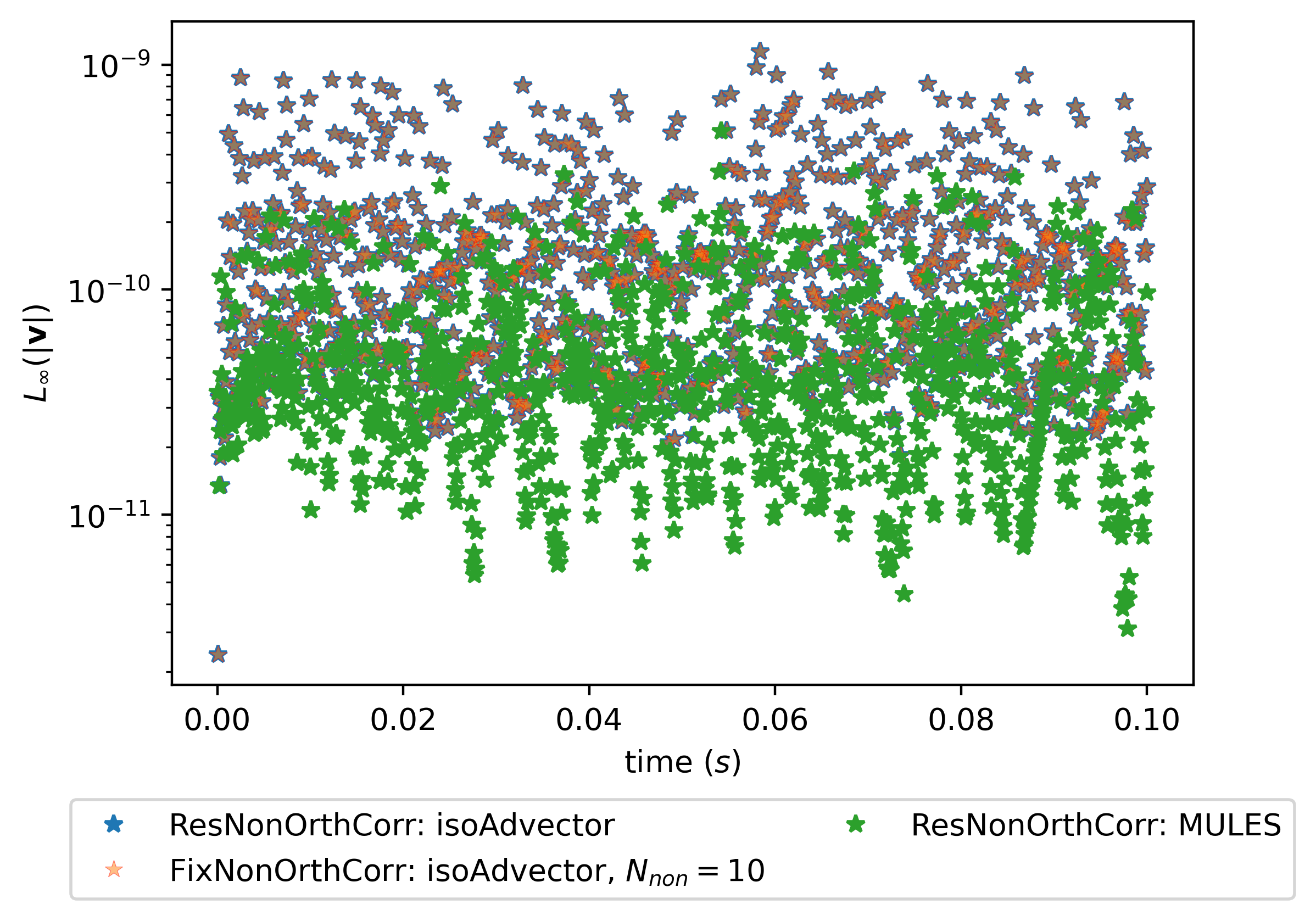}
      \caption{perturbMesh: $\frac{\Delta x}{L} = 1/60$}
     \end{subfigure}
     \begin{subfigure}{.45\textwidth}
      \centering
      \includegraphics[width=\linewidth]{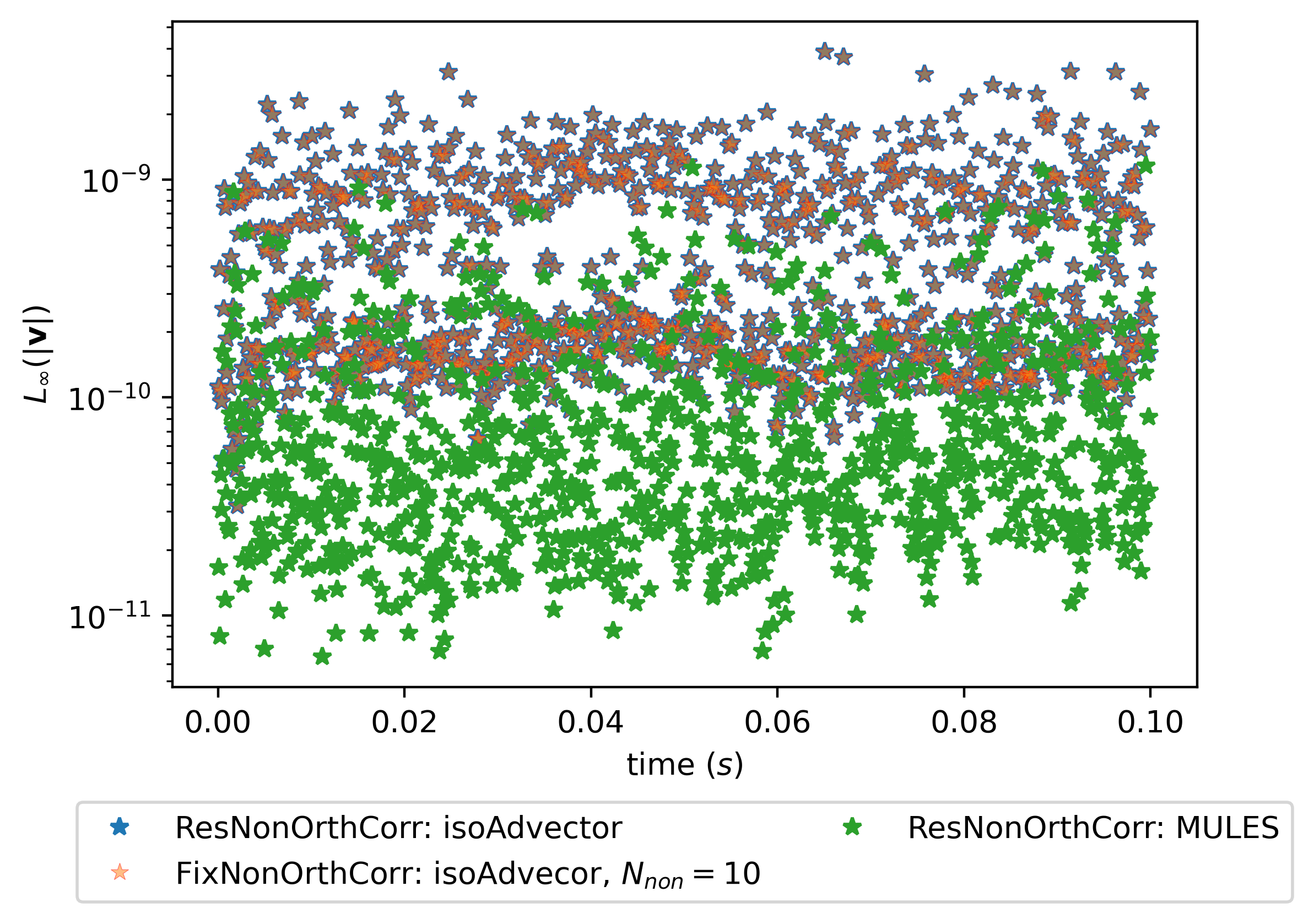}
      \caption{perturbMesh: $\frac{\Delta x}{L} = 1/90$}
     \end{subfigure}    
    \caption{ The temporal evolution of velocity error norm $L_{\infty}(|\v|)$ using MULES with ResNonOrthCorr for a stationary droplet in equilibrium on perturbMesh with different resolutions. These results show that the proposed method is directly applicable to the algebraic VOF method.}
    \label{fig:stationaryDroplet_perturbMesh_MULES}
\end{figure}

\section{Supplementary results for Stationary water column in equilibrium}
\label{Appendix:figures_waterColumn}

\begin{figure}[htb]
     \begin{subfigure}{.45\textwidth}
      \centering
      \includegraphics[width=\linewidth]{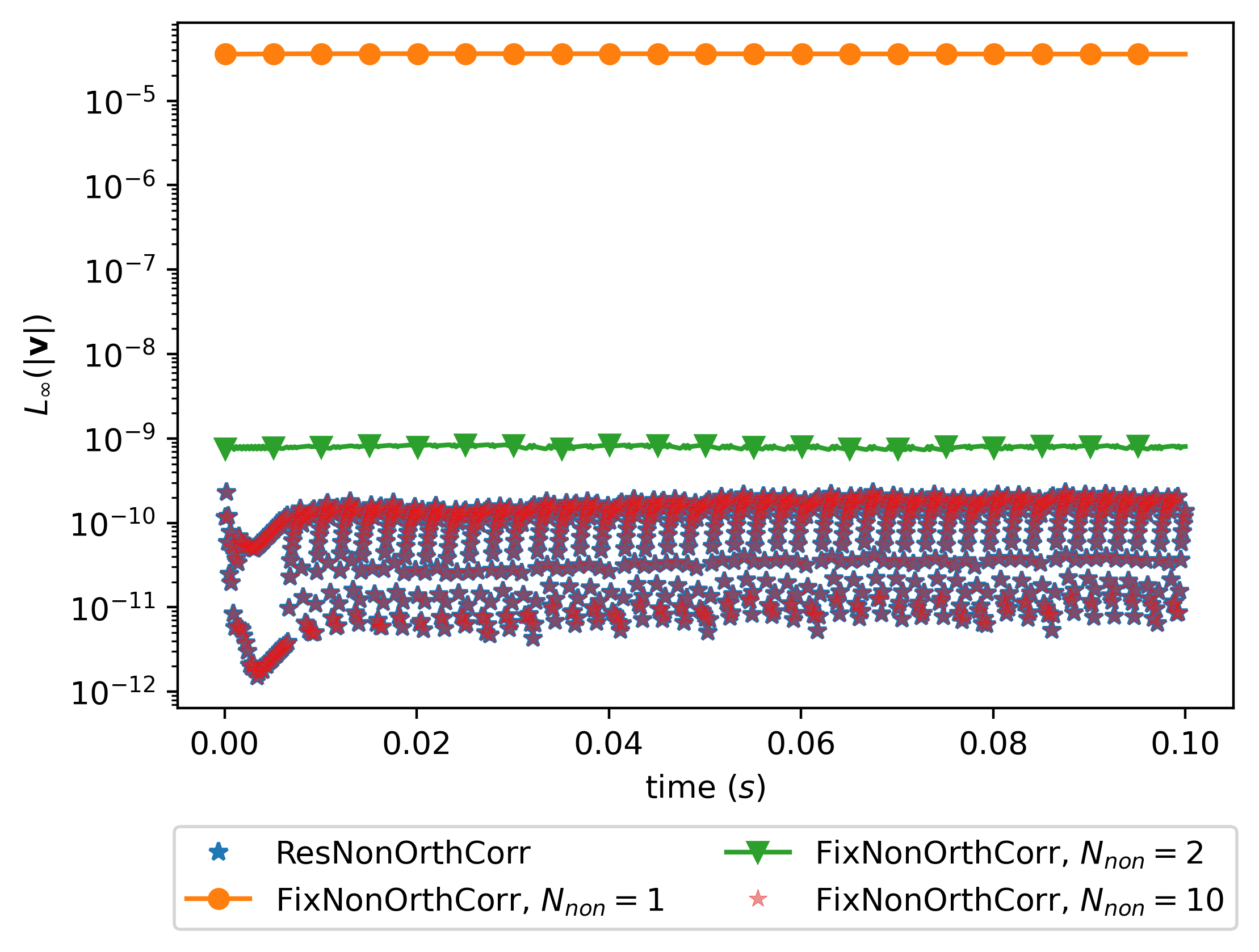}
      \caption{perturbMesh: $\frac{\Delta x}{L} = 1/30$}
     \end{subfigure}
     \begin{subfigure}{.45\textwidth}
      \centering
      \includegraphics[width=\linewidth]{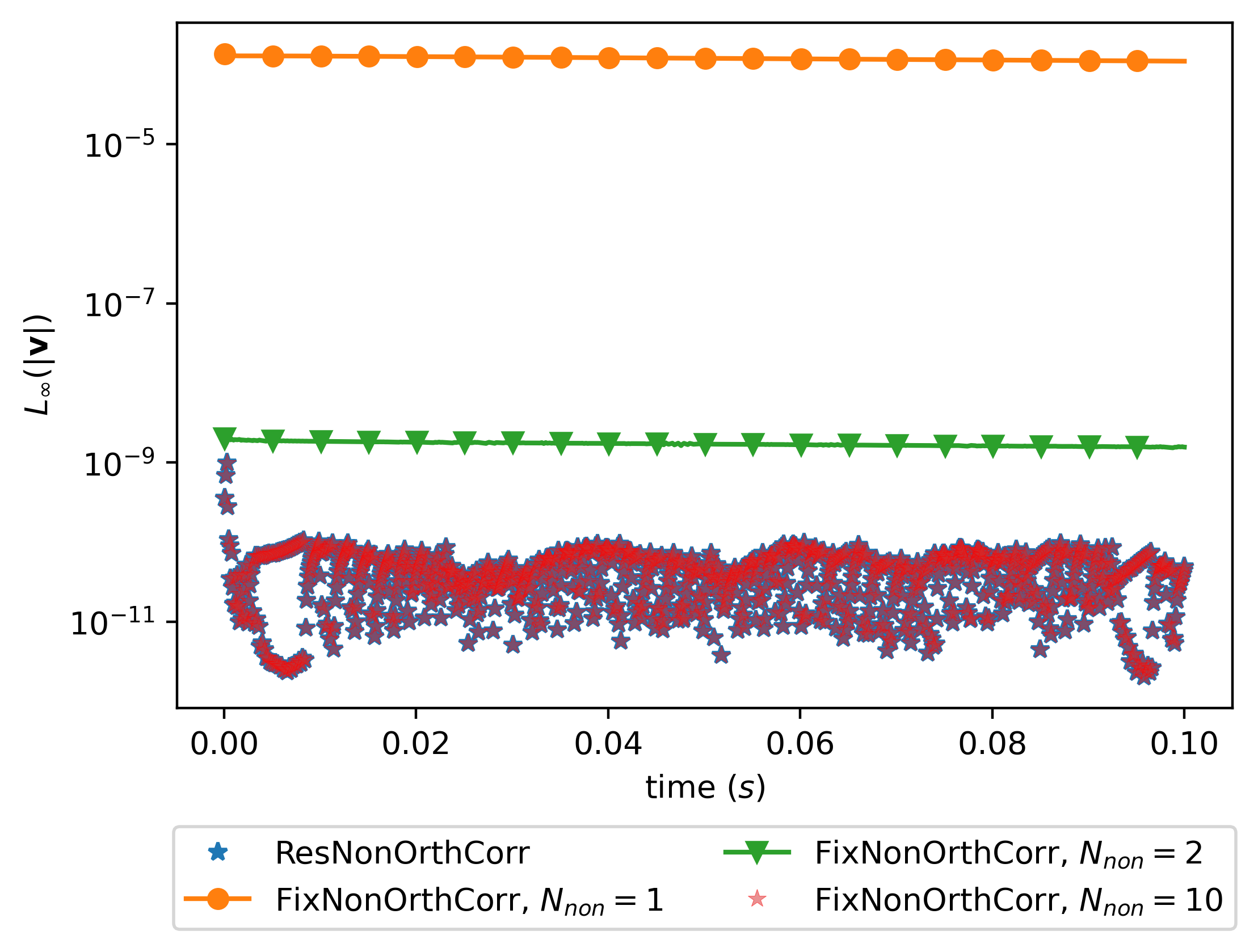}
      \caption{perturbMesh: $\frac{\Delta x}{L} = 1/60$}
     \end{subfigure}
     \begin{subfigure}{.45\textwidth}
      \centering
      \includegraphics[width=\linewidth]{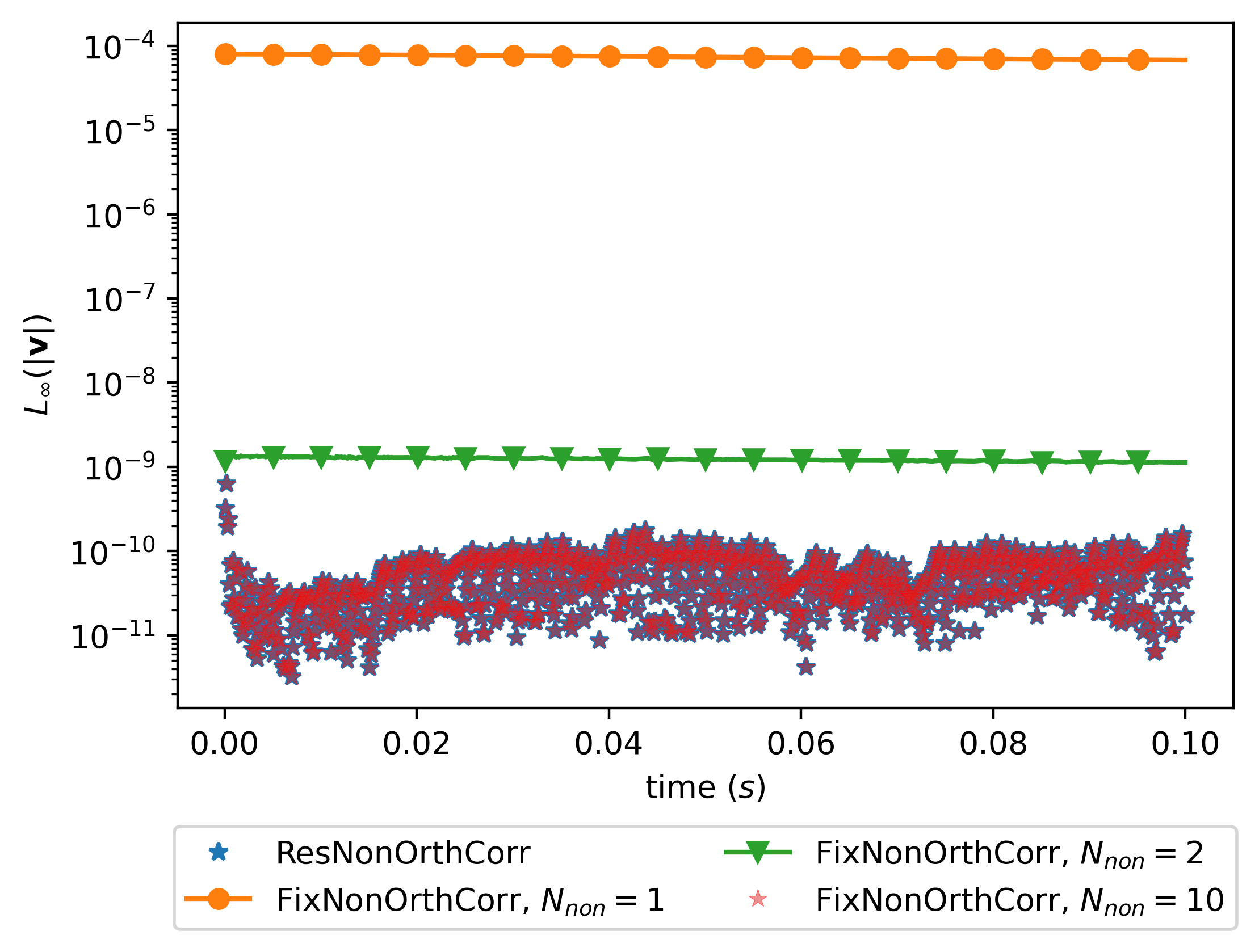}
      \caption{perturbMesh: $\frac{\Delta x}{L} = 1/90$}
     \end{subfigure}    
    \caption{The temporal evolution of velocity error norm $L_{\infty}(|\v|)$ with ResNonOrthCorr and FixNonOrthCorr ($N_{non}= [1,2,10]$) for a stationary water column in equilibrium on perturbMesh with different resolutions. Although $N_{non}=10$ achieves force-balance, it is a problem-dependent "free" parameter that the proposed method does not use.}
    \label{fig:waterColumn_perturbMesh}
\end{figure}
\begin{figure}[htb]
     \begin{subfigure}{.45\textwidth}
      \centering
      \includegraphics[width=\linewidth]{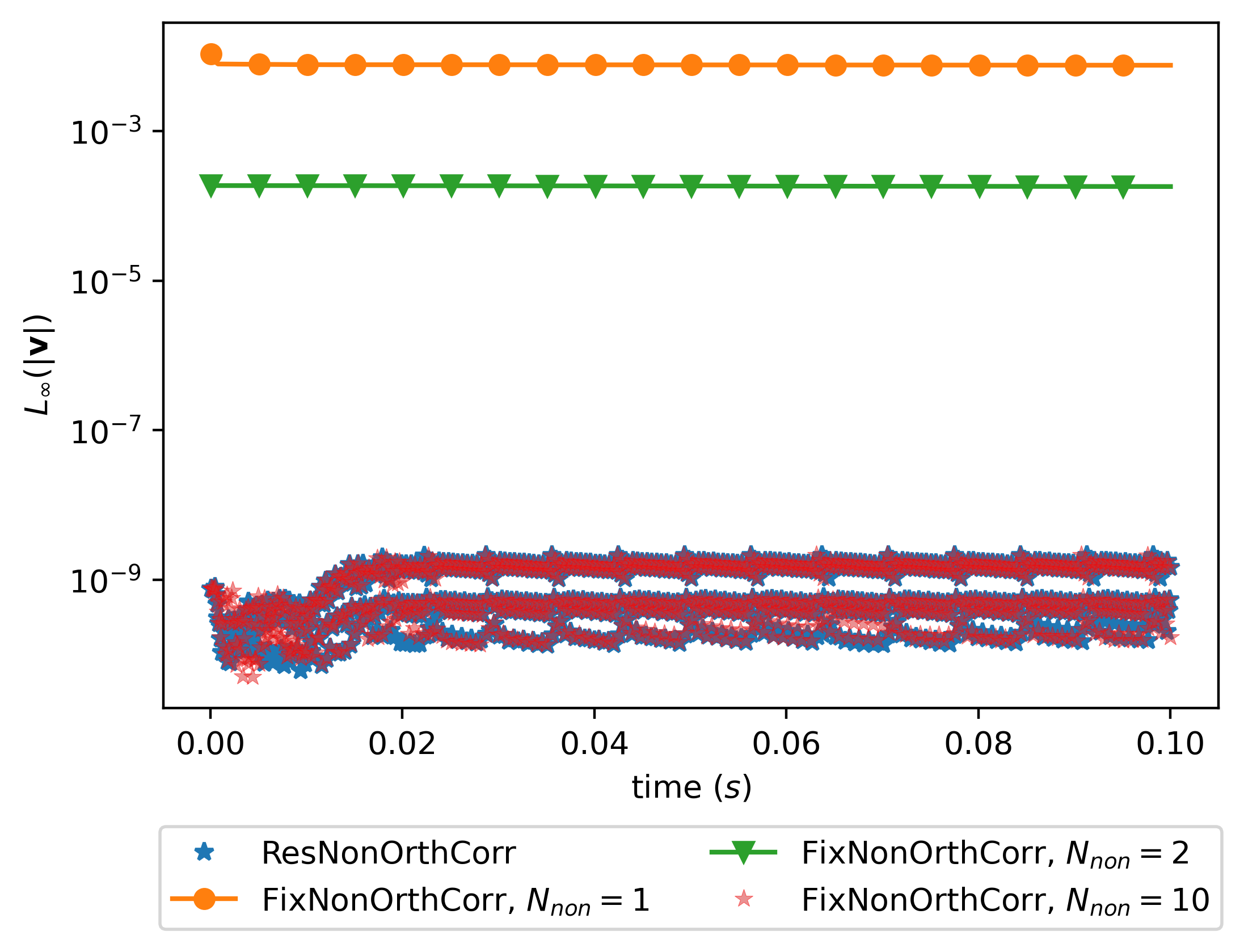}
      \caption{polyMesh: $\frac{\Delta x}{L} = 1/30$}
     \end{subfigure}
     \begin{subfigure}{.45\textwidth}
      \centering
      \includegraphics[width=\linewidth]{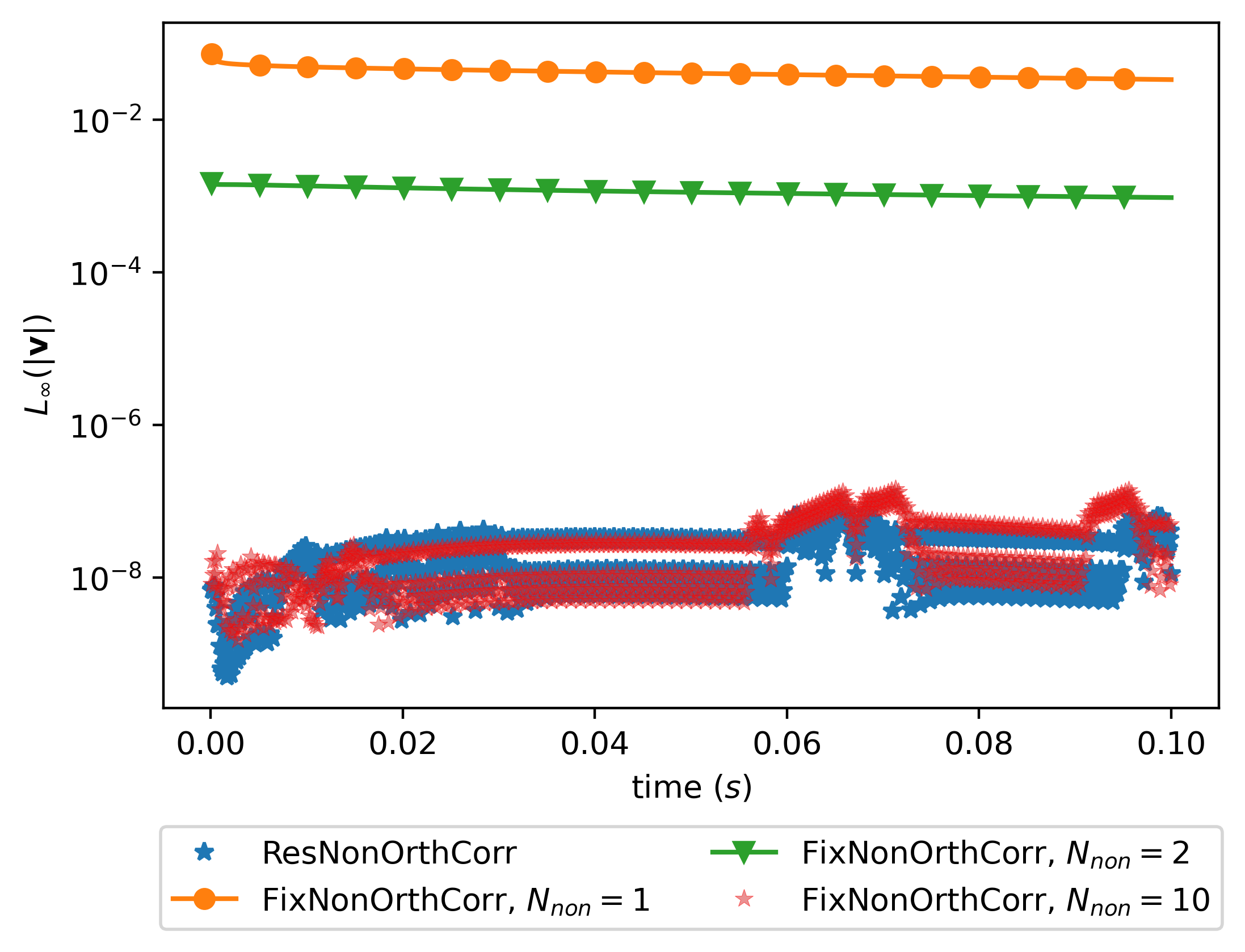}
      \caption{polyMesh: $\frac{\Delta x}{L} = 1/60$}
     \end{subfigure}
     \begin{subfigure}{.45\textwidth}
      \centering
      \includegraphics[width=\linewidth]{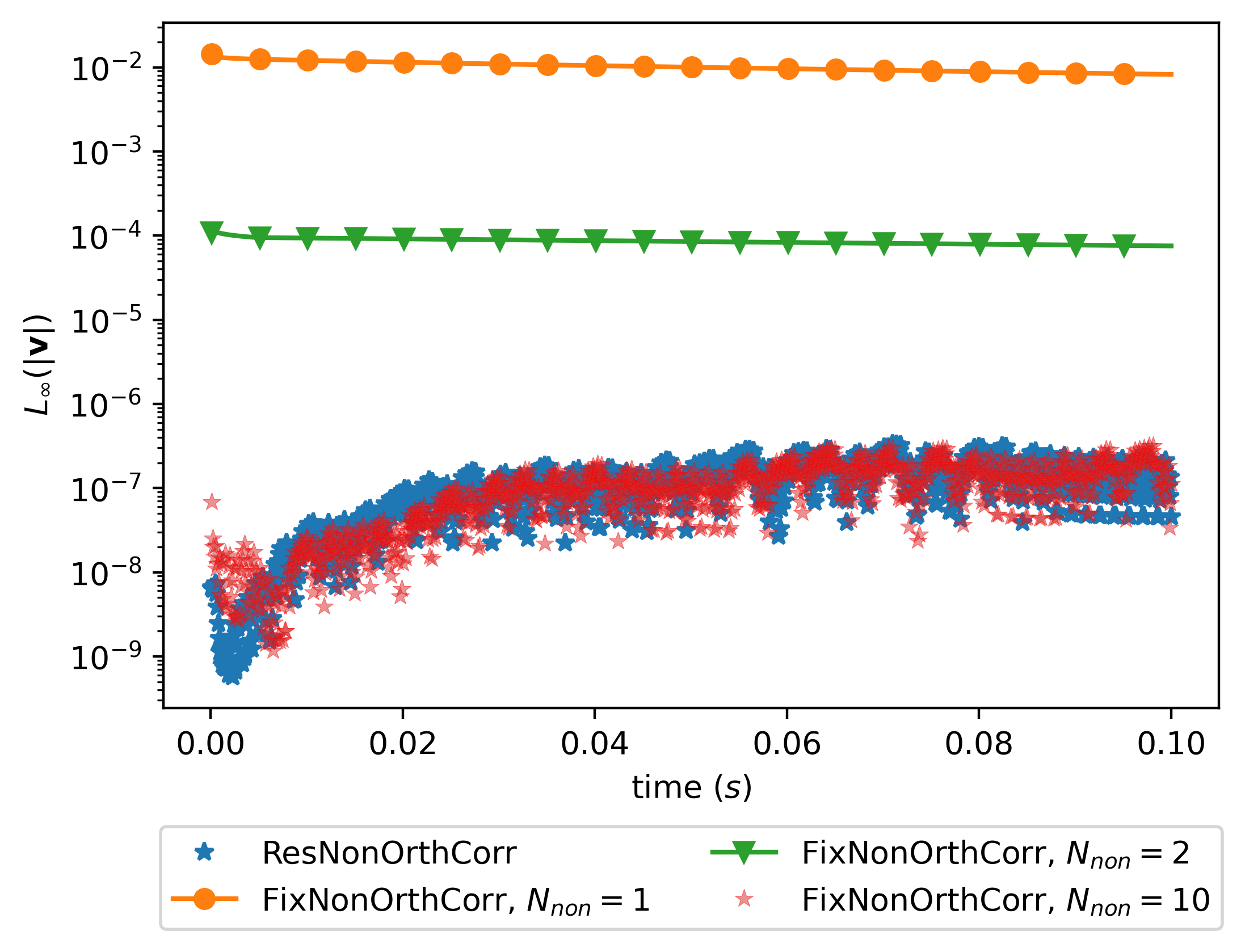}
      \caption{polyMesh: $\frac{\Delta x}{L} = 1/90$}
     \end{subfigure}    
    \caption{ The temporal evolution of velocity error norm $L_{\infty}(|\v|)$ with ResNonOrthCorr and FixNonOrthCorr ($N_{non}= [1,2,10]$) for a stationary column in equilibrium on polyMesh with different resolutions. Although the polyhedral mesh has very low non-orthogonality in the bulk, near walls, even for a cubic domain, non-orthogonality is substantial, and the force-imbalance is efficiently and effectively restored by the proposed method.}
    \label{fig:waterColumn_polyMesh}
\end{figure}


\bibliographystyle{elsarticle-num-names} 
\bibliography{literature}


\clearpage

\end{document}